\documentclass[12pt]{article}

\usepackage{graphicx,a4,epsfig,float,rotate,amsmath,amsfonts}    
\usepackage{latexsym,color,cite}
\newcommand{\be}{\begin{equation}}
\newcommand{\ee}{\end{equation}}
\newcommand{\bea}{\begin{eqnarray}}
\newcommand{\eea}{\end{eqnarray}}
\newcommand{\nnnl}{\nonumber \\}
\newcommand{\fs}{\,.}
\newcommand{\co}{\,,}
\newcommand{\scs}{\, , \,}

\newcommand{\ed}{\end{document}}
\newcommand{\bc}{\begin{center}}
\newcommand{\ec}{\end{center}}

\newcommand{\eq}{\begin{eqnarray}}
\newcommand{\en}{\end{eqnarray}}
\newcommand{\spi}{s}
\begin{document}

\thispagestyle{empty}

\begin{flushright}
{\footnotesize HISKP--TH--08/13}
\end{flushright}

\bigskip

\vspace{3cm}
\bc{\Large{\bf Isospin breaking in \boldmath $K_{l4}$ decays}}

\vspace{0.5cm}

G.~Colangelo$^a$,
J.~Gasser$^a$ and  
A.~Rusetsky$^{b}$

\vspace{2em}

\begin{tabular}{c}
$^a\,$Center for Research and Education in Fundamental Physics,\\
  Institute for Theoretical Physics,  University of Bern,\\
Sidlerstr. 5, CH-3012 Bern, Switzerland\\
$^b\,$Helmholtz--Institut f\"ur Strahlen-- und Kernphysik,\\
 Bethe Center for Theoretical Physics,\\
Universit\"at Bonn, Nussallee~14--16, D--53115 Bonn, Germany\\
\end{tabular} 

\ec

\vspace{1cm}

{\abstract{Data on $K_{e4}$ decays allow one to extract experimental
    information on the elastic $\pi\pi$ scattering amplitude near
    threshold, and to confront the outcome of the analysis with predictions
    made in the framework of QCD. These predictions concern an isospin
    symmetric world, while experiments are carried out in the real world,
    where isospin breaking effects -- generated by electromagnetic
    interactions and by the mass difference of the up and down quarks --
    are always present.  We discuss the corrections required to account for
    these, so that a meaningful comparison with the predictions becomes
    possible. In particular, we note that there is a spectacular isospin
    breaking effect in $K_{e4}$ decays. Once it is taken into account, the
    previous discrepancy between NA48/2 data on $K_{e4}$ decays and the
    prediction of $\pi\pi$ scattering lengths disappears.}}

\vskip1cm

{\footnotesize{\begin{tabular}{ll}
{\bf{Pacs:}}$\!\!\!\!$& 11.30.Rd, 12.38.Aw, 13.20.-v, 12.39.Fe\\
{\bf{Keywords:}}$\!\!\!\!$& $K_{e4}$ decays, isospin breaking,
pion-pion scattering,\\& chiral symmetries
\end{tabular}}
}
\clearpage

\tableofcontents

\section{Introduction}
Chiral perturbation theory (ChPT) \cite{weinberg79}, combined with Roy
equations, allows one to make very precise predictions for the values of
the threshold parameters in elastic $\pi\pi$ scattering \cite{scattnpb} --
for a status report, see e.g. the contribution of one of us at
KAON'07~\cite{colangelokaon07}.  Several experiments have measured the $\pi
\pi$ interaction at low energy with a precision such that it is now
possible to confront these predictions with data: i) $K^+\to \pi^+\pi^-
e^+\nu_e$ decays
\cite{ke4old,ke4NA48/2,blochkaon07,blochanacapri,talksna48}, ii) the
pionium lifetime, measured by the DIRAC collaboration \cite{DIRAC}, iii)
the cusp effect in $K\to 3\pi$ decays, investigated by the NA48/2 and KTeV
collaborations \cite{k3pNA48kTeV,talksna48}.  The experiments performed by
the NA48/2 collaboration have generated an impressive data basis, as a
result of which the matrix elements of $K\to 3\pi$ and of $K_{e4}$ decays
can be determined with an unprecedented accuracy.

The theoretical predictions and the measurements are performed in two
different settings: the predictions concern pure QCD, in the isospin
symmetry limit $m_u=m_d$, with photons absent -- a {\it paradise world}.
To be more precise, the convention is to choose the quark masses and the
renormalization group invariant scale of QCD such that the pion and the
kaon masses coincide with the values of the charged ones, and the pion
decay constant is $F_\pi=92.4$ MeV.  [We do not specify the masses of the
heavy quarks, because in the present context, their precise values do not
matter.]  On the other hand, experiments are all carried out in the
presence of isospin breaking effects, generated by real and virtual
photons, and by the mass difference of the up and down quarks: this is the
{\it real world}, described by the Standard Model.  We are thus faced with
the problem to find the relation between quantities measured in the real
world, where isospin breaking effects are always present, and the
predictions made in the paradise world.  It is the aim of the present
article to provide this relation.

In early experiments \cite{ke4old}, the effects of real and virtual photons
were estimated by considering a simplified model for the weak interactions,
and taking into account photon effects through minimal coupling, working at
lowest nontrivial order in $\alpha_\mathrm{QED}$. In analyses of NA48/2
data before spring 2007, real and virtual photon effects were treated in a
factorized manner, applying the Coulomb factor and using the program PHOTOS
\cite{photos} -- see Ref.~\cite{ke4NA48/2} for details.  It is clear that
both approaches missed the effects generated by the pion and kaon mass
differences, and by the quark mass difference $m_d-m_u$.  It turned out
that these effects are quite spectacular
\cite{internalnote,tarasovke4,gevorkyanke4,descotesanacapri}: 
when taken into account, all previous discrepancies between
data and prediction for the scattering lengths disappear
\cite{colangelokaon07,blochkaon07,blochanacapri,talksna48}\footnote{In
  fact, the relevant expressions for the pertinent corrections are
  contained already in the early works of Cuplov and Nehme \cite{cuplov},
  but they went, unfortunately, largely unnoticed in the literature.}.

The outline of the paper is as follows. In section
\ref{sec:isospinsymmetry}, we recall how $K_{e4}$ data are analysed in the
isospin symmetry limit, based on Watson's theorem, while section
\ref{sec:theframework} describes the framework in which we take isospin
breaking effects into account. In section \ref{sec:watson} we investigate
-- for the case of (Lorentz) scalar pion form factors -- the effect of
isospin breaking mass differences using unitarity and analyticity alone. We
illustrate the outcome of this investigation in section \ref{sec:examples}
with two explicit examples in the framework of Quantum Field Theory, before
we return to $K_{e4}$ decays in section \ref{sec:ke4breaking}, where we
display the result for the changes in the phase of the form factors, and
for the phase--removed form factors themselves. These results allow us in
section \ref{sec:thefits} to perform fits to $K_{e4}$ data, based on
numerical solutions to Roy equations, and to determine in this manner
experimental values for the $\pi\pi$ scattering lengths. A comparison with
related work available in the literature is provided in the following
section \ref{sec:comparison}, while a summary and concluding remarks are
given in section \ref{sec:summary}. Appendix \ref{app:threshold} contains
proofs of the general statements made on the low--energy structure of
phases and form factors in the main text. Finally, Appendix \ref{app:model}
contains material needed for the calculations of form factors in a
non-relativistic framework.

\section{$K_{e4}$ decays: isospin symmetry limit}
\label{sec:isospinsymmetry}
To set notation and to explain the manner in which  $K_{e4}$ decays allow
one to measure $\pi\pi$ phase shifts, let us consider here the decays  in
the isospin symmetry limit $m_u=m_d,\alpha_\mathrm{QED}$ =0. The matrix
element for $K^+(p)\to \pi^+(p_1)\pi^-(p_2)e^+(p_e)\nu_e(p_\nu)$ is 
\bea
      T = \frac{G_F}{\sqrt{2}} V^\star_{us} \bar{u} (p_\nu) \gamma^\mu
      (1-\gamma_5)  v (p_e) (V_\mu - A_\mu),
\eea
where the last factor denotes hadronic matrix elements of the strangeness 
changing  (vector and axial vector) currents,
\bea
V_\mu-A_\mu & = & \langle \pi^+ (p_1) \pi^- (p_2) \,\mbox{out}\mid
(\bar s\gamma_\mu u-\bar s\gamma_\mu\gamma_5 u) \mid K^+ (p)  \rangle\fs
\eea
In the following, we concentrate on the matrix element of the axial vector
current, because it carries information on the $\pi\pi$ final state
interactions and, in particular, on the $\pi\pi$ phase shifts. One
decomposes $A_\mu$ into Lorentz scalars, 
\bea
\label{eq:axial} 
A_\mu  =  -i\frac{1}{M_K} \left [ (p_1+p_2)_\mu F +
(p_1-p_2)_\mu G + (p_e+p_\nu)_\mu R \right ]\fs
\eea
The form factors $F,G, R$  are holomorphic functions of the three variables
\bea
\spi=(p_1+p_2)^2\scs t=(p_1-p)^2\scs u=(p_2-p)^2\, \fs
\eea
Sometimes, it is useful to use instead
\bea
\spi=(p_1+p_2)^2\scs s_\ell=(p_e+p_\nu)^2\scs \cos{\theta_\pi}\scs
\eea
where $\theta_\pi$ is the angle of the $\pi^+$ in the CM system of the two
charged pions, with respect to the dipion line of flight in the rest system
of the kaon \cite{cabibbomaksymovicz}.
In the isospin symmetry limit, one identifies the $\pi\pi$ phase shifts in
the matrix element in a standard manner, by performing a partial wave
expansion, and using unitarity and analyticity, although, in the present
case, this is a slightly intricate endeavour \cite{partialwaveexpansion}. 
It is useful to introduce a particular combination of form factors (we omit
isospin indices), 
\bea
F_1=F+\frac{(M_K^2-\spi-s_\ell)\sigma}{\lambda(M_K^2,\spi,s_\ell)^{1/2}} \cos{\theta_\pi}G\fs
\eea
Here, $\sigma=\sqrt{1-4M_\pi^2/\spi}$, and $\lambda(x,y,z)$ is the triangle
function. The form factor $F_1$ has a simple partial wave expansion,
\bea\label{eq:formfactorsf_i}
F_1=\sum_{k\geq 0}  P_k(\cos{\theta_\pi}) f_k(\spi,s_\ell)\fs
\eea
For fixed $s_\ell$, the amplitudes $f_k$ are holomorphic\footnote{See
  Appendix \ref{app:notation} for our notation of the various sets in the
  complex plane.} in $\mathbb{C}_{LR}(4M_\pi^2)$. 
In the elastic region, the form factors $f_k$ carry the $\pi\pi$ phase shifts 
in the pertinent isospin channel \cite{partialwaveexpansion}. For $f_0$ and
$f_1$, the relation is
\bea\label{eq:fphase}
f_0^+&=&e^{2i\delta_0}f_0^-\,,\,f_1^+=e^{2i\delta_1}f_1^-\,;\quad
s\in[4M_\pi^2,16 M_\pi^2]\scs 
\eea
where $\delta_0$ ($\delta_1$) denotes the  phase shift of the isospin zero
S-wave (isospin one P-wave), and $f_n^+ (f_n^-)$ stands for the form factor
evaluated above (below) the cut, 
\bea
f_n^\pm&=&f_n(\spi\pm i\epsilon,s_\ell)\fs
\eea
The phase--removed form factor can be Taylor expanded at threshold. 
For the quantity $f_0$, the expansion reads
\eq\label{eq:expandq2}
e^{-i\delta_0}f^+_0=c_0+c_2q^2+O(q^4)\scs\quad q^2=\frac{s}{4M_\pi^2}-1\scs
\en
with coefficients $c_i$ that depend on $s_\ell$.
Because the modulus squared $|F_1|^2$ enters the decay rate, one can
measure the phase shift difference $\delta_0-\delta_1$ in $K_{e4}$ decay
experiments. In the remaining part of this article, we investigate the
manner in which the relations Eqs.~(\ref{eq:fphase},\ref{eq:expandq2}) are
modified in the presence of isospin symmetry breaking effects, and how the
analysis of $K_{e4}$ data must be modified in order to determine the
$\pi\pi$ phase shifts. 
\begin{figure}[t]
\begin{center}
\includegraphics[width=13cm]{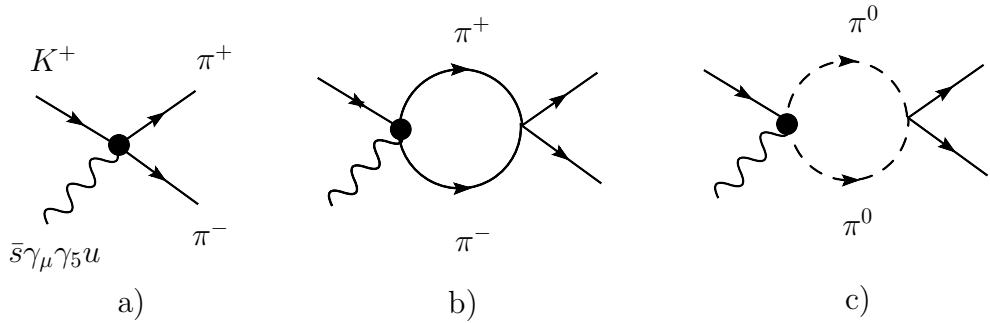}
\end{center}
\caption{Some of the graphs that contribute to the matrix element of the
  axial current at tree and one--loop order.  The filled vertex indicates
  that the axial current also couples to a single kaon line. That graph
  contributes to the form factor $R$. There are many additional graphs at
  one--loop order, not displayed in the figure.}
\label{fig:treeoneloop}
\end{figure}

Before proceeding, let us consider the form factors at one loop in chiral
perturbation theory and verify that $F_1$ indeed has the behaviour just
discussed. For this, we recall the pertinent effective Lagrangian
\bea
{\cal L}_2=\frac{F_0^2}{4}\langle D_\mu U D^\mu U^\dagger +2B_0{\cal M}(U+U^\dagger)\rangle\,,
\eea
where the covariant derivative $D_\mu U$ contains the external vector and
axial vector currents, and ${\cal M}=\mathrm{diag}(\hat m,\hat m,m_s)$.  
Some of the graphs that contribute at tree--level and at one loop are
displayed in figure \ref{fig:treeoneloop}. The full result is
\cite{kl4oneloop} 
\bea\label{eq:oneloopke4}
f_0(\spi,s_\ell)=\frac{M_K}{\sqrt{2}F_0}\left\{1+\Delta(\spi)+H(\spi,s_\ell)+O(p^4)\right\}\,,
\eea
with
\bea\label{eq:jbar}
\Delta(\spi)&=&\frac{1}{2F_0^2}(2\spi-M_\pi^2){\bar J} (\spi)\,,\nnnl
16\pi^2\bar J(\spi)&=&
\sigma\left(\ln{\frac{1-\sigma}{1+\sigma}}+i\pi\right) +2\,,
\quad \spi \geq 4M_\pi^2\fs 
\eea
Here, $M_\pi$ ($F_0$) denotes the pion mass (pion decay constant), at
leading order in the chiral expansion. The quantity $H(\spi,s_\ell)$ is
real in the interval of elastic $\pi\pi$ scattering. 
It is now seen that  $f_0$ indeed has the property (\ref{eq:fphase}) at this 
order in the low--energy expansion,
 with
\bea\label{eq:phasesymm}
\delta_0=\frac{(2\spi-M_\pi^2)}{32\pi F_0^2}\sigma\,.
\eea
This is the phase shift of the  isospin zero S-wave, in tree
approximation. The amplitude $f_1$ has a very similar structure
\cite{kl4oneloop}, containing the phase shift of the isospin one P-wave,
again in tree approximation. {\underline{Remark:}} To get the result
Eq.~\eqref{eq:phasesymm}, we have used an expansion  around the chiral
limit $m_u=m_d=m_s=0$, as a result of which the pertinent pion decay
constant $F_0$ in this limit appears. To the order considered here, we may
replace $F_0$ by the pion decay constant $F$ in the chiral limit
$m_u=m_d=0,m_s\neq 0$.  
It seems to us that this is a more natural choice when discussing
the $\pi \pi$ phase shift, and we will, therefore, use $F_0\Rightarrow F$ in 
numerical analyses in the rest of this article.

\section{Isospin breaking: the framework}
\label{sec:theframework}
The relation \eqref{eq:fphase} holds in the isospin symmetry limit.  On the
other hand, $K_{e4}$ decays happen to occur in the real world where
$m_d\neq m_u, \alpha_\mathrm{QED}\neq 0$, as a result of which the
relations Eq.~\eqref{eq:fphase} do not hold anymore.  Here, we discuss how
this situation can be modelled in view of the already published analyses of
$K_{e4}$ decays.

In early experiments \cite{ke4old}, the effects of real and virtual photons
were estimated by considering a simplified model for the weak interactions,
and taking into account photon effects through minimal coupling, working at
lowest nontrivial order in $\alpha_\mathrm{QED}$. In analyses of NA48/2
data before summer 2007, real and virtual photon effects were treated in a
factorized manner, applying the Coulomb factor and using the program PHOTOS
\cite{photos} -- see Ref.~\cite{ke4NA48/2} for details.  It is clear that
both approaches missed the effects generated by the pion and kaon mass
differences, and by the quark mass difference $m_d-m_u$.  These must thus
be taken into account separately. We assume that they can be evaluated in
factorized form as well, and write symbolically in case of the NA48/2
analysis

\vskip3mm \bc 
\mbox{\it Full isospin breaking effects = Coulomb factor
  $\times$ PHOTOS $\times$ mass effects}\nonumber \ec 
\vskip3mm As a
practical way of proceeding, which should catch the main effects, we
propose to correct also earlier analyses \cite{ke4old} with the last
factor.

\begin{figure}[t]
\begin{center}
\includegraphics[width=4cm]{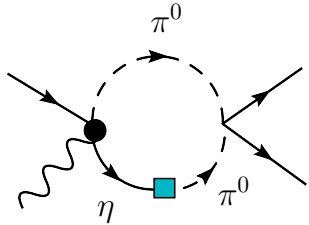}
\end{center}
\caption{One--loop graph in the $K_{e4}$ decay amplitude
that emerges due to the $\pi^0\eta$ mixing. There is no counterpart of this
graph in the scalar form factor. }
\label{fig:mixing}
\end{figure}

We now discuss the manner in which mass effects may be evaluated. As the
correction turns out to be small (although not negligible), a perturbative
method is appropriate. In the following, we use effective field theory
techniques to perform the calculation. As real and virtual photons have
already been taken into account, we need a framework that accounts for mass
effects only.  An obvious candidate is the chiral lagrangian itself, in the
absence of real photons.  This can be achieved by modifying the lagrangian
${\cal L}_2$: one adapts the quark mass matrix, ${\cal M}\to
\mathrm{diag}(m_u,m_d,m_s)$, and adds mass breaking terms of
electromagnetic origin \cite{chiralp4},
\bea\label{eq:Lbreaking}
{\cal L}_2\to {\cal L}_2+C\langle QUQU^\dagger\rangle\,,\qquad
Q=\frac{e}{3}\mathrm{diag}(2,-1,-1)\,, 
\eea
where $C$ is a  low--energy constant that breaks the isospin symmetry of 
the meson masses, such that $M_\pi\neq M_{\pi^0},M_K\neq M_{K^0}$. The
symbol $e$ stands for the electric charge. At order $p^4$, there will be
additional terms, classified in Refs.~\cite{AnnalsNPB,chiralp4}.
The effect of the replacement (\ref{eq:Lbreaking}) is twofold: first, because
the pion masses split, the loop contributions generated by the diagrams
displayed in  Fig.~\ref{fig:treeoneloop}b),c) have a different
threshold. Second, in addition to the  graphs displayed in figure
\ref{fig:treeoneloop}, there is a new contribution shown in figure
\ref{fig:mixing}: the kaon interacts with the axial current to generate a
$\pi^0\eta$ intermediate state. Because $m_u\neq m_d$, the $\eta$ can
transform back into a neutral pion, that then re--scatters with the second
neutral pion into a charged pion pair. We perform  later in this article a
quantitative analysis of these effects.

In summary, we propose the following framework to purify published phase
shifts from mass effects. 
\begin{itemize}
\item[i)]
 We assume that
the published phase shifts  correspond to the ones obtained
in a world defined by the lagrangian \eqref{eq:Lbreaking}, 
\bea\label{eq:fphasenew}
f_k^+=e^{2i\psi_k}f_k^-\,,\, k=0,1\,;\, \spi\geq 4M_\pi^2\fs
\eea
We refer in the following to $\psi_k$ as ``measured phase shifts''.
\item[ii)]
In order to get into contact with the isospin symmetric phase shifts
$\delta_k$, we note that 
\bea\label{eq:corrections}
\delta_k&=&\psi_k-(\psi_k-\delta_k)\scs k=0,1\fs
\eea
The differences $\psi_k-\delta_k$ on the right--hand side are small, and
can be calculated in the effective field theory framework outlined above.
After subtracting these from the measured phase shifts  $\psi_k$, one gets
$\delta_k$, which can then be confronted with predictions in the framework
of ChPT. 
\end{itemize}

While this procedure does not present a complete and full analysis of
radiative corrections in $K_{e4}$ decays, it allows one to purify published
phase shifts from mass effects and thus to hopefully retain the main
effects of isospin breaking. [One may envisage a more ambitious procedure
\cite{internalknecht}, by working out the relevant matrix elements in the
framework of ChPT including photons and leptons \cite{chptlept}, and then
constructing a new event generator, to be used in a new analysis of
$K_{e4}$ decays\footnote{ This would also take care of the Coulomb phase,
  whose effect is suggested to be substantial in
  Refs.~\cite{cuplov,tarasovke4,gevorkyanke4}. While the Coulomb phase acts
  in the same direction as the mass effects considered here -- it increases
  the difference $\psi_0-\psi_1-(\delta_0-\delta_1)$, see section
  \ref{sec:thefits} -- it is not clear which part thereof is already
  included in PHOTOS. We, therefore, prefer to stick to the procedure
  proposed here, because these are corrections that are definitely not
  included in PHOTOS.}. Eventually, such an analysis might lead to an
improved algorithm. However, we consider this to be a long term project.]

The framework proposed here allows one to investigate the strictures
imposed on the form factors by unitarity and analyticity, even without
relying on any specific details of the lagrangian Eq.~\eqref{eq:Lbreaking}.
We find it instructive to discuss this fact in some detail in the following
section -- before performing the explicit evaluation of isospin breaking
effects in $K_{e4}$ decays -- because it illustrates that, qualitatively,
the effects we are finding at the end in those decays do not rely on a
specific lagrangian framework, but are present in any theory that breaks
isospin symmetry and incorporates unitarity and analyticity. [On the other
hand, to work out the effects in a quantitative manner, a specific
underlying theory is needed.]  Readers who are not interested in this
general setting may wish to skip the following two sections and continue
directly with section \ref{sec:ke4breaking}, where we take up the case of
$K_{e4}$ decays again, and where we calculate the corrections
$\psi_0-\delta_0$ and $\psi_1-\delta_1$, and investigate their effect on
the $\pi\pi$ scattering lengths.

\section{Watson's theorem and all that}
\label{sec:watson}

Here, we discuss Watson's theorem and its modification in the isospin
breaking case by using analyticity and unitarity arguments alone.  As
already mentioned, it is not quite straightforward to work out unitarity
and analyticity constraints on the axial current matrix element
Eq.~\eqref{eq:axial} which is relevant here, because this matrix element
describes the scattering process $K+\mbox{axial current} \to \pi+\pi$,
which has a rather complicated structure. On the other hand, the physical
effects generated by isospin breaking interactions are also present in
simpler matrix elements like the (Lorentz) scalar form factor of the pion.
We, therefore, illustrate the basic facts in this simpler setting. It will
become obvious that the same line of reasoning could be carried over in the
same general framework to the $K_{e4}$ form factors, although with much
more labor.

To set the framework, we consider matrix elements of a hermitian Lorentz
scalar current $ j (x)$, which is taken to be isoscalar in the absence of
isospin breaking interactions. To simplify the analysis further, we
consider the case where the isospin symmetric theory corresponds to the
standard $SU(2)_R\times SU(2)_L\to SU(2)$ scenario of spontaneously broken
two-flavour QCD, with pions only.  Emission of real photons is excluded, as
a result of which all matrix elements are infrared finite.

As isospin symmetry is assumed to be broken, it is convenient to use state
vectors that are labelled by the physical pion states. The two form factors
are 
\eq\label{eq:formfac} \langle 0|j
(0)|\pi^+(p_1)\pi^-(p_2);\mbox{in}\rangle&=&-F_c(s)\, ,
\nonumber\\[2mm]
\langle 0|j (0)|\pi^0(p_1)\pi^0(p_2);\mbox{in}\rangle&=&F_0(s)\, ,
\quad\quad s=(p_1+p_2)^2\fs 
\en 
In the isospin symmetry limit, one has $F_c=F_0$ in the Condon-Shortley
phase convention used here.  These form factors are boundary values of
functions $F_k(z)$ which are assumed to be 
i) holomorphic in $\mathbb{C}_R(4M_{\pi^0}^2)$, and ii) real on the real
axis for $s < 4M_{\pi^0}^2$.  For further specifications of $F_k$ needed to
arrive at the results described below, see Appendix \ref{app:threshold}.

We also need the elastic $\pi\pi\to \pi\pi$ scattering matrix elements and
consider the following three channels 
in $\pi^a\pi^b\to\pi^c\pi^d : (ab;cd)= (00;00),(+-;00),(+-;+-)$. 
We denote the  scattering matrix elements by
$T_i(s,t)$, where $i=0,x,c$  are the channel labels in the above order,
and $s,t$ stand for the Mandelstam variables. The partial-wave expansion
reads
\eq
T_i(s,t)=32\pi\left[h_i(s)+\sum_{l=1}^\infty P_l(\cos\vartheta)
h^l_i(s)\right]\, ,
\en 
Here,  $\vartheta$ stands for the scattering angle, and $h_i(s)$ denote the pertinent 
S-waves, which are needed in the following. 
They are  boundary values of functions $h_i(z)$ which are assumed to be
  i) holomorphic in $\mathbb{C}_{LR}(4M_{\pi^0}^2)$, and 
ii) real on the real axis for  $0<s < 4M_{\pi^0}^2$.
For further specifications of $h_k$ needed to arrive at the results
described below, see Appendix \ref{app:threshold}.

\vskip3mm

The unitarity conditions for the form factors and for the partial waves 
read on the upper rim of the cut 
\eq\label{eq:unitarityF}
{\rm Im}\, F(s )&=&T(s )\rho(s)F^*(s )\scs\nnnl
{\rm Im}\, T(s )&=&T(s )\rho(s)T^*(s )\,;\,\quad 4M_{\pi^0}^2\leq s\leq16
M_{\pi^0}^2\fs 
\en
We have used the matrix notation
\eq\label{eq:matricesFT}
&&F=\left(\!\!\begin{array}{c}
F_c\\ F_0\end{array}\!\!\right)\scs\quad\quad
T=\left(\!\!\begin{array}{rr}
h_c & -h_x\\ -h_x & h_0
\end{array}\!\!\right)\scs
\nonumber\\[2mm]
&&\rho(s)=\left(\!\!\begin{array}{cc} 2\sigma(s)\,\theta(s-4M_\pi^2)&0\\0& 
\sigma_0(s)\,\theta(s-4M_{\pi^0}^2)\end{array}\!\!\right)\scs
\en
together with
\eq\label{eq:sigmas}
\sigma(s)=\sqrt{1-\frac{4M_\pi^2}{s}}\quad,\quad
\sigma_0(s)=\sqrt{1-\frac{4M_{\pi^0}^2}{s}}\, .
\en
The quantity $\sigma(z)$ is holomorphic in $\mathbb{C}_{LR}(4M_\pi^2)$. 
On the upper rim of the right--hand cut we take $\sigma(z)$ to be 
positive and real, and the analytic continuation thereof 
elsewhere in the  complex $z$-plane. 
Analogous statements hold for $\sigma_0(z)$.

\subsection{Isospin symmetry limit}
\label{subsec:watsoniso}
\begin{figure}[h]
\bc
\includegraphics[width=6.5cm]{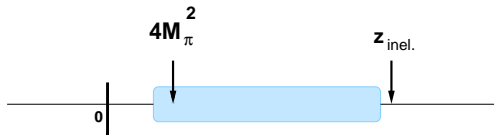}
\caption{Region of analyticity in the complex $z$-plane, isospin symmetric
  case. The phase--removed form factor $\hat F_c(z)$ is holomorphic in the
  shaded region. The symbol $z_{inel.}$ denotes the inelastic threshold
  $z=16 M_\pi^2$.}\label{fig:holsymmt}
\ec
\end{figure}

In the isospin symmetry limit, one has
\eq\label{eq:I02}
\left(h_c\scs h_x\scs h_0\right)=\frac{1}{6}\left(t_2+2t_0
\scs 2t_2-2t_0\scs 4t_2+2t_0\right)\fs
\en
The quantity $t_I$  denotes the partial wave amplitude with angular
momentum zero and isospin $I$. With $F_c=F_0$, the unitarity relations
Eq.~\eqref{eq:unitarityF} decouple, 
\bea
{\rm Im} F_c&=&t_0\sigma   F^*_c\scs\nnnl
{\rm Im} \,t_I&=&\sigma  |t_I|^2\scs I=0,2\fs
\eea
We analyse in Appendix \ref{app:threshold} the singularity structure of the
partial waves and of the form factor which follow from these relations
\cite{threshold}. The result is as follows: 
\begin{itemize}
\item[i)]

The form factor develops a square root singularity at the threshold
$z=4M_\pi^2$, 
\bea
F_c(z)=A(z)+i\sigma B(z)\scs
\eea
where $A,B$ are meromorphic in $\mathbb{C}_{LR}(16 M_\pi^2)$, and real on
the real axis in the interval $0<s<16 M_\pi^2$. 
\item[ii)]
Define the phase
\bea\label{eq:phasessym}
{\theta(z)}&=&\arctan{\frac{\sigma(z) B(z)}{A(z)}}\fs
\eea
In the elastic interval, $\theta$ coincides with the phase shift $\delta_0$
of the partial wave $t_0$ -- this is Watson's theorem. 
The above definition of $\theta$ holds also for complex values of $z$. As a
result, $\theta$ is holomorphic in the shaded region shown in
Fig.~\ref{fig:holsymmt}, cut along the positive real axis for $s>4M_\pi^2$.
[Although $\theta(z)$ is complex in general, we keep calling this quantity
{\it phase} for simplicity.]
\item[iii)]
In analogy to Eq.~\eqref{eq:expandq2}, we define the phase--removed form
factor  
\bea\label{eq:phaseremoved}
\hat F_c(z)&=&e^{-i\theta(z)}F_c(z) = 
A\left(1+\frac{\sigma^2B^2}{A^2}\right)^{1/2}\,.
\eea
$\hat F_c$ is  holomorphic in the shaded region  shown in figure
\ref{fig:holsymmt}, and real on the real axis in that region. It
can therefore be Taylor expanded at threshold, 
\bea
\hat F_c=\sum_{i\geq 0}e_i(z-4 M_\pi^2)^i\scs
\eea
with real coefficients $e_i$, cf. with Eq.~\eqref{eq:expandq2}.
\end{itemize}
Analogous facts were (implicitly and explicitly) used for the axial form
factors $f_k$ in Eq.~\eqref{eq:formfactorsf_i} in  all previous analyses of
$K_{e4}$ decays. 

\begin{figure}[h]
\bc
\includegraphics[width=6.5cm]{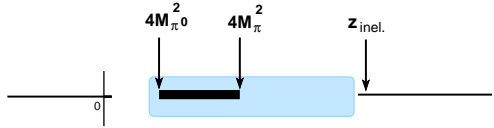}
\caption{Region of analyticity, isospin broken case. The phase--removed
  form factors $\hat F_k$ are holomorphic in the (cut) shaded region. The
  symbol $z_{inel.}$ denotes the inelastic threshold $z=16 M_{\pi^0}^2$.}
\label{fig:holbrokent}
\ec
\end{figure}
\subsection{Broken isospin symmetry}
We now consider the case where the underlying theory contains isospin
breaking interactions. To render the discussion simple, we assume that
these terms are reasonably small, such that the isospin symmetric values of
the form factors and of the scattering matrix are only slightly modified.
This can always be arranged by adjusting the relevant parameters in the
lagrangian, and allows us to discuss the changes induced in a simpler
manner.  We analyse the singularity structure of the form factors $F_{c,0}$
in Eq.~\eqref{eq:formfac} in the general case in Appendix
\ref{app:threshold} as well, and we find that the four statements i) to iv)
made above are modified in the following manner:
\begin{itemize}
\item[i')]
The form factors  have the threshold behaviour
\bea\label{eq:threshold_fisobroken}
F_k=A_k+i\sigma B_k+i\sigma_0 C_k+\sigma\sigma_0 D_k\,;\, k=c,0\scs
\eea
with coefficients $A_k,\ldots, D_k$ that are meromorphic in
$\mathbb{C}_{LR}(16 M_{\pi^0}^2)$, and  real on the real axis in the
interval $0<s<16 M_{\pi^0}^2$.  
\item[ii')]
Let
\bea\label{eq:phasesbroken}
{\theta_k(z)}&=&\arctan{\frac{\sigma(z) B_k(z)+\sigma_0(z) C_k(z)}{A_k(z)
+\sigma(z)\sigma_0 (z)D_k(z)}}\,,
\eea
and introduce the phase--removed form factors $\hat F_k$,
\bea
\hat F_k(z)&=&e^{-i\theta_k(z)}F_k(z)\fs
\eea
The quantities $\hat F_k$ are holomorphic  in the (cut) shaded region shown
in Fig. \ref{fig:holbrokent}, and real on the real axis in the intervals
$0<s<4M_{\pi^0}^2$ and $4M_\pi^2<s<16M_{\pi^0}^2$. 
\item[iii')]
In contrast to the isospin symmetric case one finds that
\begin{itemize}
\item[a)]
the $\pi\pi$ scattering amplitude does not fully determine the phases
$\theta_k$ below the inelastic region, 
\item[b)] $\theta_c$ does not vanish at the threshold $s=4 M_\pi^2\scs$
\bea
\theta_c\not\to 0\scs s\to 4M_\pi^2\fs
\eea
\end{itemize}
\item[iv')]
Again in contrast to the isospin symmetric case,
the form factor $\hat F_c$ develops a square root singularity at $z=4M_\pi^2$,
\bea\label{eq:1qq2}
\hat F_c(s)=\bar e_0+\bar e_1q +O(q^2)\scs s\searrow 4M_\pi^2\,;\, 
\eea
with real coefficients $\bar e_i$ .
\end{itemize}
This result shows that the phase--removed form factor $\hat F_c$ cannot be
expanded in a Taylor series at the threshold $s=4M_\pi^2$. [One may be
tempted to get rid of the problem by defining a form factor where the
pertinent Omn\`es factor is removed. In the present case, this is not what
one wants to do, because the aim is to measure the phase shift, 
which is needed to evaluate the Omn\`es factor.]

\section{Explicit examples}\label{sec:examples}
We find it useful to provide  in this section examples of form factors
that illustrate the analytic properties worked out  above.

\subsection{Scalar form factor in  a non-relativistic effective theory}
\label{sec:scalarff}

In order to describe the behaviour of the scalar form factor in the
low--energy region $q^2=s/(4M_\pi^2)-1\ll 1,~\Delta_\pi =
M_\pi^2-M_{\pi^0}^2\ll M_\pi^2$, we use the framework of non-relativistic
effective field theory (NREFT) as developed in
Refs.~\cite{cuspwe,cuspwe0,cusprad,physrep}. This formulation is especially
convenient to study the singularity structure near threshold, because the
locations of the low--energy singularities coincide with that in the
relativistic QFT to all orders in the low--energy expansion.  A similar
non-relativistic framework has recently been used to study $K\to 3\pi$
decays in the vicinity of cusps~\cite{cuspwe,cuspwe0,cusprad}. The effects
that we are addressing in this section have the same physical origin as the
cusps in $K\to 3\pi$ decays. They emerge, because the final state
interactions involve both, $\pi^+\pi^-$ and $\pi^0\pi^0$ pairs, with
different masses $M_\pi\neq M_{\pi^0}$. We display the pertinent
non-relativistic lagrangian in Appendix \ref{app:model}.

\begin{figure}[t]
\begin{center}
\includegraphics[width=13cm]{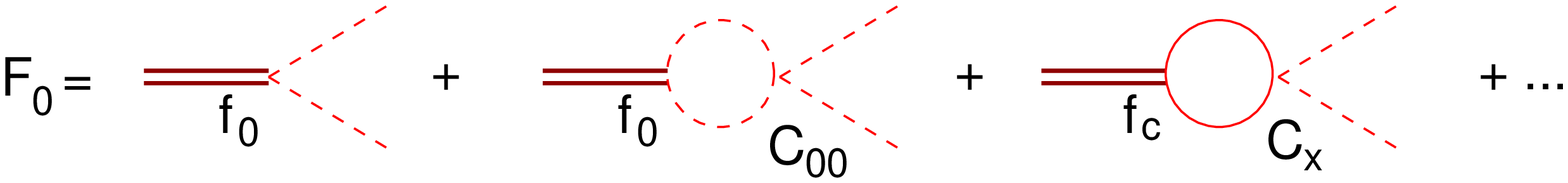}
\end{center}
\caption{Bubble diagrams, contributing to the neutral scalar form factor in
  the non-relativistic effective theory.
Solid and dashed lines correspond to  charged and neutral pions,
respectively.}
\label{fig:bubblesum}
\end{figure}

The scalar form factors $F_c(s)$ and $F_0(s)$, defined in
Eq.~(\ref{eq:defscalar}), are given by a sum of bubble diagrams,
see Fig.~\ref{fig:bubblesum}. The infinite series of these bubble diagrams
can be explicitly re-summed. The resulting expressions for $F_c(s)$,
$F_0(s)$ are boundary values of  meromorphic functions $F_c(z)$, $F_0(z)$
at $z\to s+i0$, with 
\eq\label{eq:explicitF}
F_c(z)&=&\frac{f_c(1-2i\sigma_0d_{0})-2i\sigma_0 f_0d_x}
{1-2i\sigma_0 d_{0}-4i\sigma d_{c}-2\sigma\sigma_0\chi}\, ,
\nonumber\\[2mm]
F_0(z)&=&\frac{-4i\sigma f_c d_x+f_0(1-4i\sigma d_{c})}
{1-2i\sigma_0 d_{0}-4i\sigma d_{c}-2\sigma\sigma_0\chi}\, ,
\en
where $d_i,~i=c,x,0$ and $f_k,~k=c,0$ denote  polynomials of first order in
$z$, and $\chi=4(d_{c}d_{0}-d_x^2)$. The coefficients of these polynomials
are expressed through  various non-relativistic couplings which, in turn,
are determined by performing the matching to the underlying relativistic
theory (see, e.g.~\cite{cuspwe,cuspwe0,cusprad,physrep}). In particular,
the polynomials $d_i$ contain 4-pion non-relativistic couplings and are 
expressed in terms of the effective-range expansion parameters for the
$\pi\pi$ scattering through the matching of the $\pi\pi$ amplitudes at
threshold. Similarly, the polynomials $f_k$ are determined from the
matching to the relativistic form factors, expanded at threshold.

Further, the quantities $d_i,f_k$ contain isospin-breaking corrections.
 We write $d_i=\bar d_i+d_i'$ (and similarly for $f_k$), where $\bar d_i$
stands for $d_i$ calculated in the isospin limit. In analogy with
Eq.~(\ref{eq:I02}), $\bar d_i$ can be expressed through two first-order
polynomials $v_0,v_2$, corresponding to  total isospin $I=0,2\scs$
\eq\label{eq:iso1}
(\bar d_c,\,\bar d_x,\,\bar
d_0)=\frac{1}{12}\,(v_2+2v_0,\,2v_2-2v_0,\,4v_2+2v_0)\, . 
\en
Here,
\eq\label{eq:iso2}
v_I(z)=a_I+\frac{z-4M_\pi^2}{4M_\pi^2}\,a_Ir_I\, ,\quad\quad I=0,2\, ,
\en
and $a_I$/$a_Ir_I$ stand for the scattering length/effective range parameter
in the isospin limit, see Ref.~\cite{cuspwe}.

Providing similar explicit expressions for the isospin-breaking corrections
in $d_i$ and $f_k$ is not possible in general. The form of these
corrections is not universal and the calculations should be performed
within a particular underlying relativistic theory. An example of
calculations in ChPT at one loop is considered in the following subsection.

\begin{figure}[t]
\begin{center}
 \includegraphics[width=14cm]{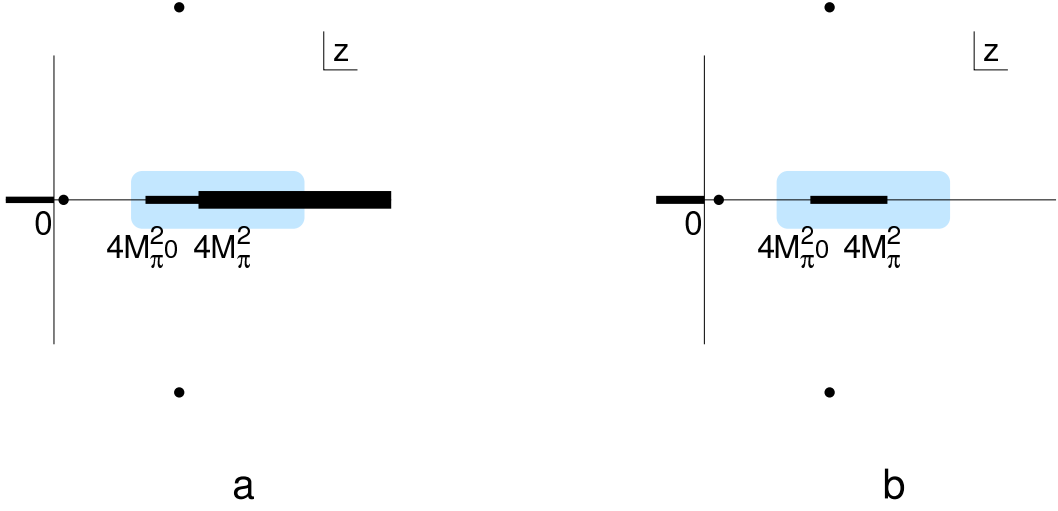}
\end{center}
 \caption{The analyticity domain of the form factor $F_c(z)$ (left panel)
and the phase--removed form factor $\hat F_c(z)$ (right panel). The shaded
area denotes the non-relativistic region and the filled circles
indicate the positions of the poles on the first Riemann sheet. Both the
left-hand cut and the poles lie  outside the non-relativistic domain.}
\label{fig:splane}
\end{figure}

We add  several comments concerning the structure of the two form factors.
\begin{itemize}
\item[i)] As already mentioned, the form factors $F_c(z)$, $F_0(z)$ are
  meromorphic functions in the cut plane displayed in
  Fig.~\ref{fig:splane}a.  The non-relativistic domain in the complex plane
  is defined as a strip surrounding the physical cut -- it is indicated
  with the shaded region in the figure.  It extends slightly below and
  above of the physical branch point, well below the first inelastic
  threshold. The maximal distance to the branch point in the
  non-relativistic region is set by the mass scale $M_\pi$.  NREFT makes
  sense only in the non-relativistic region -- the cut along the negative
  $z$-axis, as well as the distant poles on the first sheet should be
  regarded as artefacts of the non-relativistic treatment.
\item[ii)] The form factors have the representation
  (\ref{eq:threshold_fisobroken}).  Constructing the $\pi\pi$ amplitudes in
  the same manner as the form factors, one finds that the unitarity
  relation given in Eq.~(\ref{eq:unitarityF}) indeed is satisfied.
\item[iii)] In order to obtain the phase--removed form factor, we define
  the phase as in Eq.~(\ref{eq:phasesbroken}), 
\eq\label{eq:deltac}
  \theta_c(z)&=&\arctan\frac{P-Q}{1+PQ}\, ,
  \nonumber\\[2mm]
  P&=&\frac{2\sigma_0 d_{0}+4\sigma d_{c}}{1-2\sigma\sigma_0\chi}\, ,
  \quad\quad Q=2\sigma_0\biggl(d_{0}+\frac{f_0}{f_c}\,d_x\biggr)\, .  
\en
  The phase--removed form factor \eq \hat
  F_c(z)=\frac{f_c}{1-2\sigma\sigma_0\chi}\,
  \biggl(\frac{1+Q^2}{1+P^2}\biggr)^{1/2} \en is holomorphic in the (cut)
  shaded region shown in Fig.~\ref{fig:splane}b.  This differs from the
  isospin symmetric case, where the cut from $4M_{\pi^0}^2$ to $4 M_\pi^2$
  is absent.

In conformity with Eq.~(\ref{eq:1qq2}), the expansion of  $\hat F_c(z)$
at threshold contains even and odd powers of the variable $q$,
 \bea\label{eq:linear}
\hat F_c(z)&=& \hat F_c(4M_\pi^2)(1+\bar e_1 q+\bar e_2q^2+O(q^3))\, ,\, s\!\! \searrow 4M_\pi^2\scs
 \nonumber\\[3mm]
\bar e_1&=&-8d_x^{\,2}M_\pi^{-1}\Delta_\pi^{1/2}+O(\Delta_\pi^{3/2})\, ,
\eea
where  $d_x$ is evaluated at the threshold $z=4M_\pi^2$.
 Using Eqs.~(\ref{eq:iso1}) and (\ref{eq:iso2}), we  get
\eq
\bar e_1=
-\frac{2\Delta_\pi^{1/2}}{9M_\pi}\,(a_0-a_2)^2\,+O(\Delta_\pi^{3/2})\simeq
-0.004\, . 
\en
At lowest order, this term  is generated by the two--loop diagram displayed
in Fig.~\ref{fig:2loopsigmac}. Numerically, the effect is tiny.

  \begin{figure}[t]
\begin{center}
  \includegraphics[width=5cm]{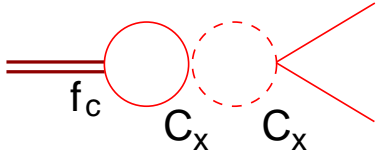}
\end{center}
  \caption{The two--loop diagram producing the linear term $\sim \sigma$ 
in the phase--removed form factor, see Eq.~(\ref{eq:linear}).}
\label{fig:2loopsigmac}
 \end{figure}

 \item[iv)]
Expanding the phase in powers of $\sigma,\sigma_0$ gives
\eq\label{eq:phaseexp}
\theta_c(z)=4\sigma d_{c}-2\sigma_0 d_x\frac{f_0}{f_c}
+O(\{\sigma,\sigma_0\}^3)\, .
\en
This expansion is very useful in the context of ChPT, since the generic 
expansion parameter here is $\{\sigma,\sigma_0\}\times d_{c,x,0}$,
with $d_{c,x,0}=O(p^2)$.
Thus, the neglected terms in Eq.~(\ref{eq:phaseexp}) are
of order $p^6$ and higher. Furthermore, in order to calculate the phase at
$O(p^4)$ (two loops), the ratio $f_0/f_c$ should be evaluated
at $O(p^2)$ and the matching of the polynomials $d_c,d_x$ should be
performed at $O(p^4)$ (at this order, $d_i(z)$ are polynomials of  second
order in $z$). To this end, it suffices to use the one--loop result both 
for the form factor and for the $\pi\pi$ scattering amplitudes.

\item[v)] As mentioned earlier, the phase $\theta_c$ is not determined by
  the $\pi\pi$ scattering amplitude alone. Indeed, $\theta_c$ depends on
  the ratio $f_0/f_c$, which can take any value in the absence of isospin
  symmetry. In addition, this ratio contains low--energy constants (LECs)
  which do not occur in the $\pi\pi$ scattering amplitude, as will be
  illustrated in subsection~\ref{sec:1loop}.

 \begin{figure}[t]
\begin{center}
 \includegraphics[width=10cm]{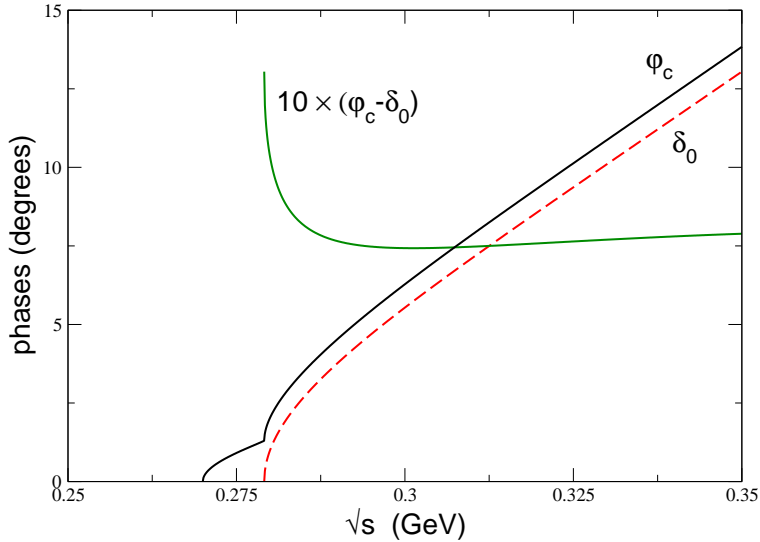}
\end{center}
 \caption{The phases $\varphi_c$ and $\delta_0$, as well as their difference.
The cusp in the phase $\varphi_c$ at the charged threshold $s=4M_\pi^2$
is clearly visible. The isospin-symmetric parts $\bar d_i$ have been determined
from the matching condition, Eqs. (\ref{eq:iso1}) and (\ref{eq:iso2}),
with the scattering
lengths and effective radii taken from Ref.~\cite{scattnpb}. 
The isospin breaking corrections $d_i'$ were evaluated in ChPT at $O(p^2)$,
see Eq. (\ref{eq:treematching}). The ratio $f_0/f_c$ was set to one.}
\label{fig:formphase}
 \end{figure}

\item[vi)]
The standard way to define the phase of the form factor
 on the real axis in the interval
$s \geq 4M_{\pi^0}^2$ is 
 \eq
 \varphi_c=\arctan\frac{\mbox{Im}\,F_c(s+i0)}{\mbox{Re}\,F_c(s+i0)}\, .
 \en
 The quantity $\varphi_c$ does not vanish at the charged threshold
 $s=4M_\pi^2$ and has a {\em cusp} there. This is illustrated in 
 Fig.~\ref{fig:formphase}, where the phase $\varphi_c$ is plotted. For
 comparison, the isospin-symmetric phase $\delta_0$
and  the difference $\varphi_c-\delta_0$ are displayed as well.
 The phases $\theta_c$ and $\varphi_c$ are identical on the real axis 
above the charged threshold,
\eq
\varphi_c(s)=\theta_c(s)\, ,\quad\quad s\geq 4M_\pi^2\, .
\en
On the other hand, they differ  e.g. in the interval $s\in
[4M_{\pi^0}^2,4M_\pi^2]$, because $\theta_c(s)$ becomes complex there.

\end{itemize}

\subsection{Form factor of pion: chiral expansion}
\label{sec:1loop}

We calculate the scalar form factor of the pion in the $SU(2)$ version of ChPT.
The Lagrangian is given by~\cite{chiralp4}
\eq
{\cal L}={\cal L}_2+{\cal L}_4\, ,
\en
where 
\eq
{\cal L}_2=\frac{F^2}{4}\langle\partial_\mu U\partial^\mu U^\dagger
+2Bs(U+U^\dagger)\rangle+C\langle QUQU^\dagger\rangle\, .
\en
In the above expression we use standard notations. In particular, $F$
denotes the pion decay constant in the chiral limit, $U$ is the pion field
matrix, the quantity $B$ is related to the quark condensate, $s={\cal
  M}+s'$ is the scalar source, ${\cal M}$ and $Q$ stand for the quark mass
and charge matrices in the $SU(2)$ case, respectively, and the coupling $C$
is proportional to $\Delta_\pi $ at lowest order in the chiral expansion.
In accordance to the discussion in section~\ref{sec:theframework}, we do
not include virtual photons.

We do not display the order $p^4$ Lagrangian ${\cal L}_4$ explicitly.  It
is given, e.g., in Refs.~\cite{chiralp4}.  Note that, since we do not
include virtual photons, the ultraviolet divergences of the
``electromagnetic'' LECs $k_i$ from Refs.~\cite{chiralp4} change in an
obvious manner.

In the calculations, the scalar current $j(x)=(2B)^{-1}(\bar u(x)u(x)
+\bar d(x)d(x))$ was used.
The scalar form factors of the pion at one loop  can be calculated in a
standard manner and are given by the following expressions,
\eq
F_c(s)&=&1+\frac{1}{2F^2}\,\biggl\{(s+4\,\Delta_\pi)\bar J(s)
+(s-M_{\pi^0}^2)\bar J_0(s)\biggr\}
\nonumber\\[2mm]
&+&\frac{1}{32\pi^2F^2}\,\biggl\{-(s+4\,\Delta_\pi)(L+1)
-(s-M_{\pi^0}^2)(L_0+1)+M_{\pi^0}^2L_0\biggr\}
\nonumber\\[2mm]
&+&\frac{4M_{\pi^0}^2}{F^2}\,l_3^r+\frac{s}{F^2}\,l_4^r+e^2{\cal K}_c^r +O(p^4)\, ,\nonumber
\en
\eq\label{eq:FcF0}
F_0(s)&=&1+\frac{1}{2F^2}\,\biggl\{2(s-M_{\pi^0}^2)\bar J(s)
+M_{\pi^0}^2\bar J_0(s)\biggr\}
\nonumber\\[2mm]
&+&\frac{1}{32\pi^2F^2}\,\biggl\{-2(s-M_{\pi^0}^2)(L+1)
-M_{\pi^0}^2(L_0+1)+2M_\pi^2L-M_{\pi^0}^2L_0\biggr\}
\nonumber\\[2mm]
&+&\frac{4M_{\pi^0}^2}{F^2}\,l_3^r+\frac{s}{F^2}\,l_4^r+e^2{\cal K}_0^r
+O(p^4)\, . 
\en
Here
\eq
L=\ln\frac{M_\pi^2}{\mu^2}\, ,\quad\quad
L_0=\ln\frac{M_{\pi^0}^2}{\mu^2}\, ,
\en
and the loop function $\bar J(s)$ is displayed in Eq.~\eqref{eq:jbar}
 [$\bar J_0$ is obtained from it by the replacement $\sigma\to\sigma_0$].
Further,
\eq\label{eq:KK}
{\cal
  K}_c^r&=&-\frac{20}{9}\,(k_1^r+k_2^r)+\frac{4}{9}\,(5k_5^r+23k_6^r+k_7^r)
+8k_8^r\, ,
\nonumber\\[2mm]
{\cal K}_0^r&=&-\frac{20}{9}\,(k_1^r+k_2^r)-2(-2k_3^r+k_4^r)
+\frac{4}{9}\,(5k_5^r+5k_6^r+k_7^r)\, .
\en
In these expressions $l_i^r$ and $k_i^r$ denote scale-dependent
renormalized LECs, the $k_i^r$ adapted to the framework considered here (no
photon loops). The quantity $\mu$ is the scale of dimensional
regularization in ChPT.

The scalar form factor of the pion in the presence of isospin breaking have
been evaluated in Ref.~\cite{Kubis:1999db}. Our expressions agree with those
of Ref.~\cite{Kubis:1999db} up to  contributions of virtual photons and
 terms of order $e^4$, $e^2(m_d-m_u)$ and $(m_d-m_u)^2$. 

The phase of the charged form factor at one loop, extracted from
Eq.~(\ref{eq:FcF0}), is given by
\bea\label{eq:phaseform}
\theta_c(s)=\frac{1}{32\pi F^2}\left\{(4\Delta_\pi +s)\sigma 
+(s-M_{\pi^0}^2)\sigma_0\right\}\, .
\eea
It is seen that $\theta_c$ indeed does not vanish at the threshold
$s=4M_\pi^2$.

Next we determine the non-relativistic couplings $d_i,f_k$, performing the
matching to ChPT at $O(p^2)$. We start from $d_i$, which can be directly read
off from the tree--level S-wave $\pi\pi$ scattering amplitudes:
\eq\label{eq:treematching}
d_c(s)&=&\frac{s+4\Delta_\pi}{128\pi F^2}+O(p^4)\scs\nnnl
d_x(s)&=&-\frac{s-M_{\pi^0}^2}{64\pi F^2}+O(p^4)\scs\nnnl
d_0(s)&=&\frac{M_{\pi^0}^2}{64\pi F^2}+O(p^4)\fs
\en
The quantities $f_k$ at $O(p^2)$ can be determined by performing the matching 
to the one--loop expressions for the form factors, given in Eq.~(\ref{eq:FcF0}).
For the ratio $f_0/f_c$ one gets
\eq
\frac{f_0}{f_c}&=&H_0+H_1(s-4M_\pi^2)+O((s-4M_\pi^2)^2)\, ,\quad\quad
\nonumber\\[2mm]
H_0&=&1+\frac{3\Delta_\pi L}{16\pi^2F^2}
+e^2({\cal K}_0^r-{\cal K}_c^r)+O(\Delta_\pi^2,p^4)\, ,
\nonumber\\[2mm]
H_1&=&\frac{\Delta_\pi}{12\pi^2F^2M_\pi^2}+O(\Delta_\pi^2,p^2)\, .
\en
We note that in the case of isospin violation the ratio
$f_0/f_c$ is not determined only by the LECs that enter the $\pi\pi$ 
amplitude. For example, even in the chiral limit 
the ratio  $f_0/f_c$ contains the LEC $k_8$, which is absent from the  
$\pi\pi$ amplitudes in this limit.   
Consequently, in the case of broken isospin,
the phase of the form factors is not determined by the $\pi\pi$ scattering 
amplitudes alone.

It is easy to check that the expression for the phase Eq.~(\ref{eq:phaseform})
can be directly obtained from Eq.~(\ref{eq:phaseexp}) by using the
result of the tree--level matching, given in Eq.~(\ref{eq:treematching}).
At this order, one may take $f_0/f_c=1+O(p^2)$.

In order to check the convergence of the chiral expansion, we have
evaluated the phase of the charged form factor at two loops. We do not
display the final result here. As expected, the next-to-leading correction
lies within approximately $30 \%$ of the leading-order result, see section
\ref{sec:thefits} for a more quantitative result.

\section{$K_{e4}$ decays: isospin breaking}
\label{sec:ke4breaking}

We now come back to the analysis of $K_{e4}$ decays. We simply need to
repeat the above analysis, this time for the matrix element
Eq.~\eqref{eq:axial} of the axial current.

Working out the contributions from diagrams Figs.~\ref{fig:treeoneloop} and
\ref{fig:mixing}, one finds that the phase $\psi_0$ of the form factor
$f_0$ [see Eq.~\eqref{eq:fphasenew}] becomes in the elastic region
$4M_\pi^2<\spi < 16 M_{\pi^0}^2$
\bea\label{eq:phase}
 \psi_0=\frac{1}{32\pi F^2}\left\{(4\Delta_\pi +\spi)\sigma 
+(\spi-M_{\pi^0}^2)\left(1+\frac{3}{2R}\right)\sigma_0\right\}+O(p^4)\, ,
\eea
with
\bea
R=\frac{m_s-\hat m}{m_d-m_u}\co
\eea
and $F$ the pion decay constant in the $SU(2)$ chiral limit.
The phase of the form factor $f_1$ does not contain at this order any
isospin breaking effects [these are generated by $\pi^0\pi^0$ intermediate
states, which cannot couple to P-waves], as a result of which one has
\bea
\psi_1=\delta_1
\eea
at this order in the low--energy expansion.
 The one--loop expressions for the form factors $F,G$
given in Refs.~\cite{cuplov}  contain the effects considered here,
up to terms of order $e^2(m_d-m_u)$. 
 \begin{figure}[h]
\begin{center}
\epsfig{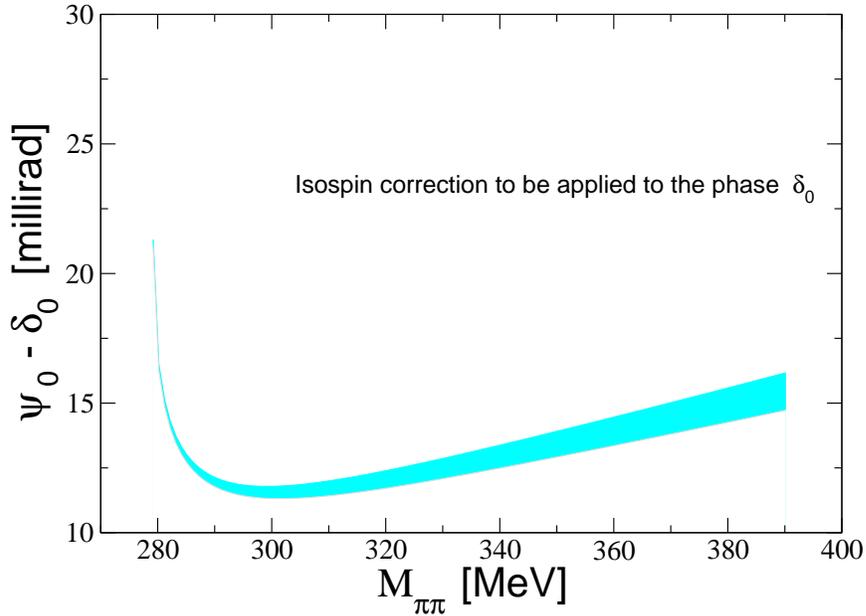}
\end{center}
\caption{The isospin breaking correction that must be subtracted from the
  phase shift $\delta$ measured in $K_{e4}$ decays. The width of the band
  reflects the uncertainty in the ratio $R=37\pm 5$.
 \label{fig:correction}}
\end{figure}

We comment on this result. First, in comparison to the pion form factor
discussed above, there is an additional term present, proportional to
$m_d-m_u$.  This is in agreement with the statement made before: the phase
is not fixed by the $\pi\pi$ amplitude alone, which does not contain any
$m_d-m_u$ terms at one--loop order.  Second, the difference to the isospin
symmetric phase now grows with $\spi$ even at one--loop order,
\bea
\psi_0-\delta_0&=&\frac{3\spi}{64\pi F^2}\frac{1}{R}+O(1)\,,\quad
\spi/M_\pi^2 \gg 1\fs 
\eea
According to the discussion in section \ref{sec:theframework}, one has to 
subtract the difference $\psi_0-\delta_0$ from the measured phase shift
before a comparison with the chiral prediction can be performed. 
In figure \ref{fig:correction} we display this difference  
in the relevant decay region, for $R=37\pm 5$.
The width of the band reflects the uncertainty in $R$. 
It is seen that the isospin correction  is quite substantial -- 
well above the  systematic and statistical uncertainties quoted for the
measured phase shift \cite{blochanacapri}.

In addition, similarly to the case of the scalar form factor, the
phase--removed form factor for the $K_{e4}$ decays is not holomorphic at
the charged threshold and, consequently, in the expansion of this form
factor even and odd powers of the variable $q$ are present, in contrast to
the isospin symmetric case, see Eq.~\eqref{eq:expandq2}.  If the value of
the form factor at $q=0$ is factored out, the coefficient of the term
linear in $q$ is the same as in the case of the scalar form factor at
leading order in ChPT. On the other hand, this is a two--loop effect and
therefore suppressed in magnitude. To obtain a reliable estimate of its
size is beyond the scope of this work.

\section{$\pi\pi$ scattering lengths from $K_{e4}$ decays}
\label{sec:thefits}
The first large-statistics experiment to measure the phase $\psi_0-\psi_1$
in $K_{e4}$ decays has been performed by the Geneva-Saclay collaboration
about thirty years ago. In this experiment about 30'000 $K_{e4}$ decays
have been collected and analysed \cite{ke4old}. More recently the E865
experiment at Brookhaven \cite{ke4old} and the NA48/2 experiment at CERN
\cite{ke4NA48/2} have each collected more than ten times the statistics and
have made possible a precise extraction of the scattering length $a_0^0$.
So precise in fact, that a proper treatment of the isospin breaking
corrections as discussed in this paper becomes essential.

In this section we discuss how the isospin breaking corrections influence
the extraction of the scattering length, and will do this for all three
data sets. We perform two kinds of analyses: we will either leave both
S-wave scattering lengths free and fit them to the data, or use a
low--energy theorem which relates both of them to the scalar radius of the
pion and end up with a one--parameter fit
\cite{Colangelo:2001sp}.
\begin{figure}[tb]
\begin{center}
\includegraphics[width=12cm]{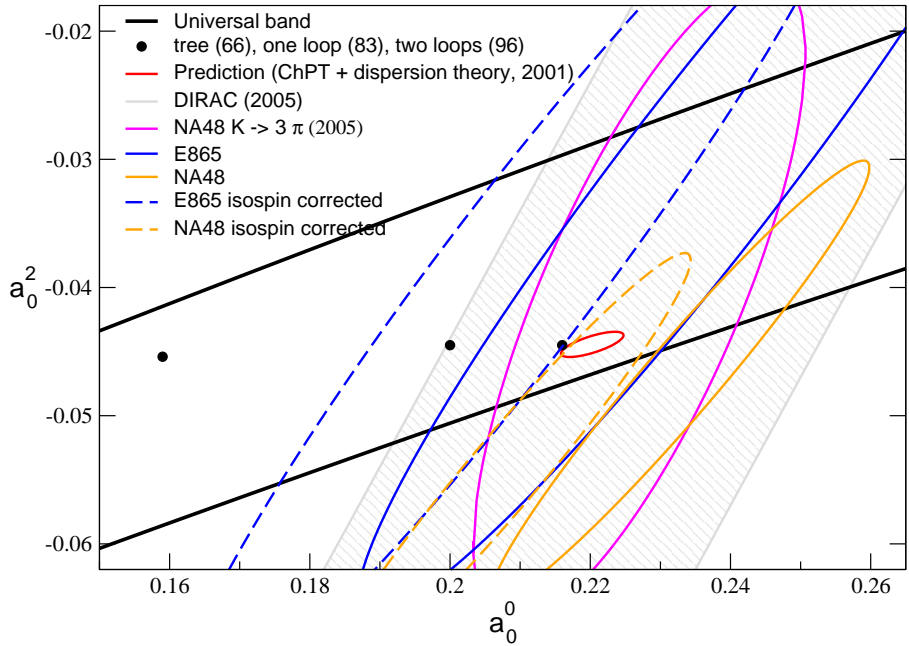}
\caption{\label{fig:a0a2} S-wave scattering lengths plane. Contours shown
  correspond to one sigma. The ellipses do not take into account any
  theoretical uncertainty.}
\end{center}
\end{figure}

For the two--parameter fits we have used the parametrization of the $\pi
\pi$ phase shifts corresponding to solutions of the Roy equations as
functions of the two S-wave scattering lengths which was provided in
\cite{Ananthanarayan:2000ht}. To evaluate numerically the isospin breaking
correction (\ref{eq:phase}) we have used $R=37 \pm 5$ and $F=86.2 \pm 0.5$
MeV \cite{Colangelo:2003hf} -- moreover we use the fact that $\psi_1=\delta_1$ at one--loop order. The
results of these fits are shown in Fig.~\ref{fig:a0a2} as one--sigma
contours (68\% probability, {\em i.e.} $\chi^2=\chi^2_\mathrm{min}+2.31$)
for the two most recent data sets, with and 
without taking into account isospin-breaking corrections.  The ellipses
corresponding to the Geneva-Saclay data are too large in comparison and for
this reason we have not shown them in the same figure.  The figure shows
that while the ellipse corresponding to the uncorrected NA48 data does not 
overlap with the theoretical prediction, after applying the isospin
breaking correction, the two overlap completely. For the ellipses
corresponding to the E865 data the situation is reversed: before applying
isospin breaking corrections there is a perfect overlap, whereas after
applying them the experimental one--sigma contour and the theory prediction
barely touch.  The figure also clearly shows that the two sets of very high
statistics data are not in very good mutual agreement: the two one--sigma
ellipses (with or without isospin breaking corrections) do not have any
significant overlap.  It has been recently pointed out by Brigitte
Bloch-Devaux that the origin of this tension lies in the data point in the
last bin of the E865 data set \cite{blochanacapri}, for which the
evaluation of the barycenter may need to be revised \cite{blochprivate}: we
have verified that indeed it is enough to remove this point from the fit to
obtain an almost perfect overlap between the two ellipses.

We take into account the constraint of the pion scalar radius as follows:
we use the numerical estimate $\langle r^2\rangle_s=0.61 \pm 0.04$
fm$^2$ \cite{scattnpb}, which implies the relation
\bea
a_0^2&=&f(\Delta a_0^0)\pm 0.0008\scs \qquad \Delta a_0^0\equiv a_0^0-0.22
\scs \nonumber \\  
f(x) &=&-0.0444 +0.236 \, x-0.61\,x^2 -9.9\, x^3\scs
\eea
where the error accounts for the various sources of uncertainty in the
input used in the Roy equation solutions. In our fits we minimize the
following $\chi^2$:
\be
\chi^2_\mathrm{tot}=\left(\frac{a_0^2-f(\Delta
a_0^0)}{0.0008}\right)^2+\chi^2_{K_{e4}} \; \; .
\ee
Since the $K_{e4}$ data are very little sensitive to
$a_0^2$, the minimum of the $\chi^2$ lies then always on the
$f(\Delta a_0^0)$ line, but in this way we also take into account the
uncertainty in the scalar radius relation. The results of the fits
are:\\[5mm]  
without applying isospin breaking corrections
\be
a_0^0 = \left\{ \begin{array}{lllll}
0.243 \pm 0.037 && \chi^2=2.2 & \mbox{Geneva-Saclay}&\mbox{[ $\,$5 data pts.]} \\
0.218 \pm 0.013 && \chi^2=5.8 & \mbox{E865} & \mbox{[ $\,$6 data pts.]} \\
0.245 \pm 0.007 && \chi^2=9.6 & \mbox{NA48} & \mbox{[10 data pts.]} \scs\\
\end{array} \right.
\ee
and after applying isospin breaking corrections
\be
a_0^0 = \left\{ \begin{array}{llll}
0.222 \pm 0.040 && \chi^2=2.1 & \mbox{Geneva-Saclay} \\
0.195 \pm 0.013 && \chi^2=6.6 & \mbox{E865} \\
0.223 \pm 0.007 && \chi^2=11.5 & \mbox{NA48} \fs\\
\end{array} \right.
\ee
Averaging the latter three independent determinations yields
$a_0^0=0.217 \pm 0.008$ where the error has been inflated by an $S=1.3$
factor according to the PDG prescription (the $\chi^2$ of the average of
the three fits is equal to 6.3). Repeating the same procedure after having
removed the last data point in the E865 set yields $a_0^0=0.221\pm0.007$
with $\chi^2\simeq 0.7$, so confirming the observation of Brigitte
Bloch-Devaux \cite{blochanacapri}. 
We also briefly comment on the vertex corrections at order $\alpha_{QED}$
mentioned in section \ref{sec:theframework}. They lower the value of
$a_0^0$ by $\simeq 0.013$. For reasons already mentioned, we do not take
these corrections into account in the present work.

The extraction of the scattering length $a_0^0$ from the phase shift
$\psi_0-\psi_1$ measured in $K_{e4}$ data requires theory input in two
steps: the first is the subject of this article, the evaluation of the
isospin breaking corrections which lead from $\psi_0-\psi_1$ to
$\delta_0-\delta_1$, and the second is the step from the phase
$\delta_0-\delta_1$ to the scattering length. Both steps can be made quite
accurately, but both are subject to a theory uncertainty which, though
small, has to be properly accounted for at the level of precision currently
reached by experiments. As far as step one is concerned, the evaluation of
the isospin breaking correction given in Eq.~(\ref{eq:phase}) relies on an
estimate of $F$ and $R$ as discussed above. A one--sigma
change in $F$ has no visible effect in the fitted value of $a_0^0$, whereas
a change of $R$ by five units shifts $a_0^0$ by $0.001$. Our one--loop
estimate of these effects is subject to higher--order corrections. We have
evaluated these in the case of the scalar form factor of the pion discussed
in previous sections. If we use the latter
as our estimate of higher--order isospin breaking effects and repeat the
fits, we get a value of $a_0^0$ shifted by $0.004$. An alternative quick
way to estimate two--loop effects is to substitute $F$ with the physical
$F_\pi$ -- doing this the fitted value of the scattering length 
changes by 0.003. We conclude that $0.004$ is a reasonable estimate of the
uncertainty due to higher--order isospin breaking corrections.
As for step two, the main source of uncertainty in the calculation of how
the Roy equation solution depend on $a_0^0$ is represented by the value of
the $\delta_0$ phase at 0.8 GeV \cite{Ananthanarayan:2000ht}, which had
been estimated to be $(82.3 \pm 3.4)^\circ$. Changing the latter by one
sigma and repeating the fits leads to a shift of $a_0^0$ by 0.004. If we
sum in squares these four different sources of uncertainty (namely: value
of $F$, value of $R$, higher--order isospin breaking corrections and value
of $\delta_0(0.8 \mbox{GeV})$ we end up with
\be
a_0^0=0.217 \pm 0.008_\mathrm{exp} \pm 0.006_\mathrm{th} \; \; ,
\ee
which represents the current best experimental determination of the isospin
zero S-wave scattering length coming from $K_{e4}$ decays. NA48 is currently
analysing more statistics \cite{blochanacapri}, and we expect that when the
updated analysis will become available, the experimental error will reach
or even become smaller than the theoretical one. Other competitive
determinations coming from the cusp in $K\to 3 \pi$ decays
\cite{k3pNA48kTeV} and from pionium lifetime \cite{DIRAC} are also expected
to be improved in the near future.

\section{Comparison with other work}
\label{sec:comparison}
The isospin-breaking corrections to the $K_{e4}$ decays have been evaluated in
the past by using various theoretical settings. We briefly review 
some recent articles where the problem has been addressed.

In Ref.~\cite{cuplov}, a full set of isospin-breaking corrections to the
$K_{e4}$ decays of charged kaons has been calculated at one loop in ChPT.
Our result for the elastic phase shift given in Eq.~(\ref{eq:phase}) agrees
with the result of Refs.~\cite{cuplov} in the absence of virtual photon
contributions. As already mentioned in section~\ref{sec:theframework},
these calculations might provide a basis for an algorithm which removes
isospin-breaking corrections from the raw experimental data.

In Refs.~\cite{tarasovke4,gevorkyanke4}, isospin-breaking corrections to
$K_{e4}$ decays have been treated in an approach which may be considered a
merger between $K$-matrix theory and a conventional quantum-mechanical
framework used to study Coulomb interactions in the final state. Although
some of the expressions in Refs.~\cite{tarasovke4,gevorkyanke4} closely
resemble the pertinent formulae of the present work, the framework used is
incomplete and, as far as we can see, incorrect.  To mention two examples,
we note that a relation between the amplitude considered there and the
$K_{e4}$ decay matrix element is not provided, and the effect caused by the
quark mass difference $m_d-m_u$ is not discussed at all. Moreover,
according to the explicit formulae given in Refs.~\cite{gevorkyanke4}, the
effect of isospin breaking -- in the absence of Coulomb interactions -- can
be expressed through the S-wave $\pi\pi$ scattering phases (in the isospin
limit) and the pion mass difference. As we show in the present article,
this is not correct. The authors of Refs.~\cite{tarasovke4,gevorkyanke4}
also provide a set of electromagnetic corrections which include both,
vertex corrections and rescattering of the virtual pions through the
exchange of Coulomb photons (the latter is a two--loop effect in our
terminology).  As already mentioned, it is not clear which part of these
corrections was already taken into account in present analyses. We
conclude that one should not use the results of
Refs.~\cite{tarasovke4,gevorkyanke4} in data analyses of $K_{e4}$ decays. 

\section{Summary and conclusions}
\label{sec:summary}

Data on $K_{e4}$ decays allow one to measure the difference of the S- and
P-wave phases $\psi_0-\psi_1$ of a particular form factor in the matrix
element of the strangeness-changing axial current. If isospin symmetry were
exact, $\psi_0-\psi_1$ would coincide with the difference of the S- and
P-wave $\pi\pi$ scattering phase shifts $\delta_0-\delta_1$ (Watson's
theorem). We have shown that the situation changes if isospin symmetry is
broken. In particular,
\begin{itemize}
\item[i)]
 The phases $\psi_0$ and $\delta_0$ are not equal. Moreover, $\psi_0$ is 
not
determined by the $\pi\pi$ scattering amplitudes alone (e.g., the former
contains the LECs which are absent in the latter).
\item[ii)]
The phase $\psi_0$ does not vanish at the threshold $s=4M_\pi^2$.
\item[iii)]
At threshold, the expansion in Eq.~\eqref{eq:expandq2}, 
which is valid in the isospin symmetry limit, becomes
$e^{-i\psi_0}f_0(s,s_\ell) =\bar c_0+\bar c_1q+\bar c_2q^2+O(q^3)\scs$
where the coefficients $\bar c_i$ depend on the variable $s_\ell$, 
and similarly for the P-wave form factor $f_1$ (although at higher order in
the chiral expansion).
\end{itemize}

Finally, in order to determine the isospin symmetric phases from data, one
uses Eq.~\eqref{eq:corrections} and calculates the difference
$\psi_k-\delta_k$, which must then be subtracted from the measured phases
$\psi_k$.  We have performed this calculation at one loop in ChPT for
$K_{e4}$ decays and have shown that the correction is substantial, well
above the statistical and systematic errors in present NA48/2 data. Once
this is taken into account, the determination of the isospin zero S-wave
scattering length from $K_{e4}$ data yields 
\[
a_0^0=0.217 \pm 0.008_\mathrm{exp} \pm 0.006_\mathrm{th} \; \; , 
\]
in excellent agreement with the chiral prediction.

\bigskip

{\bf Acknowledgments}
\begin{sloppypar}
  We are grateful to the late F.J. Yndur\'ain for useful comments at an
  early stage of this work.  We thank all participants of the $K_{e4}$
  workshop in Bern (March 2007) for interesting discussions, which have
  inspired the present investigations.  We thank B. Ananthanarayan,
  B.~Bloch--Devaux, I. Caprini, S. Gevorkyan, Ch. Hanhart, Ulf-G. Mei\ss
  ner, B.~Kubis, H.~Leutwyler, J. Pel\'aez, and Z. Was for most enjoyable
  discussions, and A.V.~Tarasov for communications concerning topics
  considered here. We are grateful to B.~Bloch-Devaux for useful comments
  concerning the manuscript, and for performing extensive fits which
  cross-checked our results for the scattering lengths.  The Center for
  Research and Education in Fundamental Physics is supported by the
  ``Innovations- und Kooperationsprojekt C-13'' of the ``Schweizerische
  Universit\"atskonferenz SUK/CRUS''.  Partial financial support under the
  EU Integrated Infrastructure Initiative Hadron Physics Project (contract
  number RII3--CT--2004--506078) and DFG (SFB/TR 16, ``Subnuclear Structure
  of Matter'') is gratefully acknowledged. This work was supported by the
  Swiss National Science Foundation, and by EU MRTN--CT--2006--035482
  (FLAVIA{\it net}).
\end{sloppypar}

\renewcommand{\thefigure}{\thesection.\arabic{figure}}
\renewcommand{\thetable}{\thesection.\arabic{table}}
\renewcommand{\theequation}{\thesection.\arabic{equation}}

\appendix

\setcounter{equation}{0}
\setcounter{figure}{0}
\setcounter{table}{0}

\newpage

\section{Notation}\label{app:notation}
 In the main text, we use the following notation for cut complex planes:
 \bea
 \mathbb{C}_{LR}(a)&=&\left\{z|z\in \mathbb{C},z\not\in(-\infty,0],z\not\in
   [a,\infty)\right\}\scs\nnnl 
 \mathbb{C}_{R}(a)&=&\left\{z|z\in \mathbb{C},z\not\in
   [a,\infty)\right\}\fs
 \eea
 \begin{itemize}
 \item[--]
 The region $\mathbb{C}_{LR}(a)$ denotes the complex plane, cut along the
 negative real axis, as well as along the positive real axis for $z\geq a$.
 \item[--]
 The region $\mathbb{C}_{R}(a)$ denotes the complex plane, 
 cut along the positive real axis for $z\geq a$.
 \end{itemize}

\section{Threshold behaviour of phases and form factors}\label{app:threshold}
We prove some of the statements made in section \ref{sec:watson} and start
with the isospin symmetric case considered in subsection
\ref{subsec:watsoniso}. The issue was discussed in the literature long ago
\cite{threshold} -- we keep the presentation therefore short.

\subsection{Isospin symmetric case}
\begin{figure}[h]
\hspace{-1mm}\includegraphics[width=6.5cm]{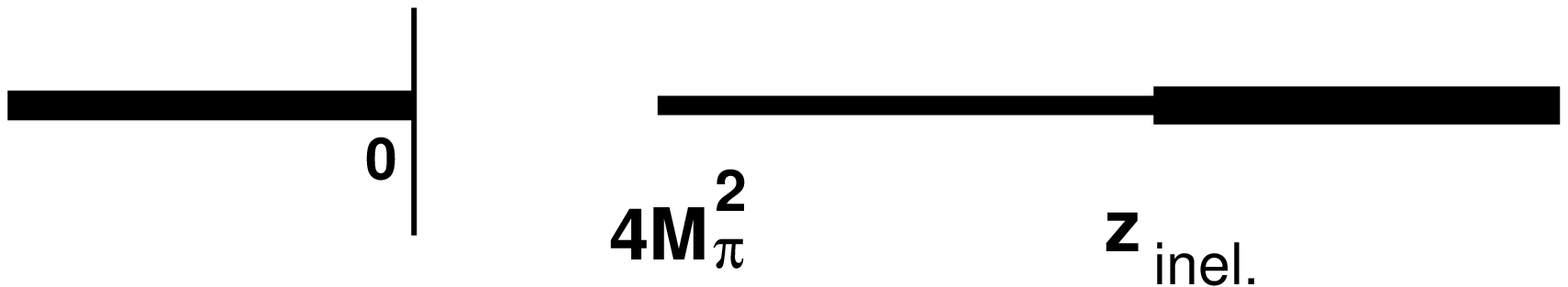} 

\vskip5mm
\hspace{2cm}\framebox{cut plane $C_I$}
\vskip-2.3cm\hspace{7cm}
\includegraphics[width=6.5cm]{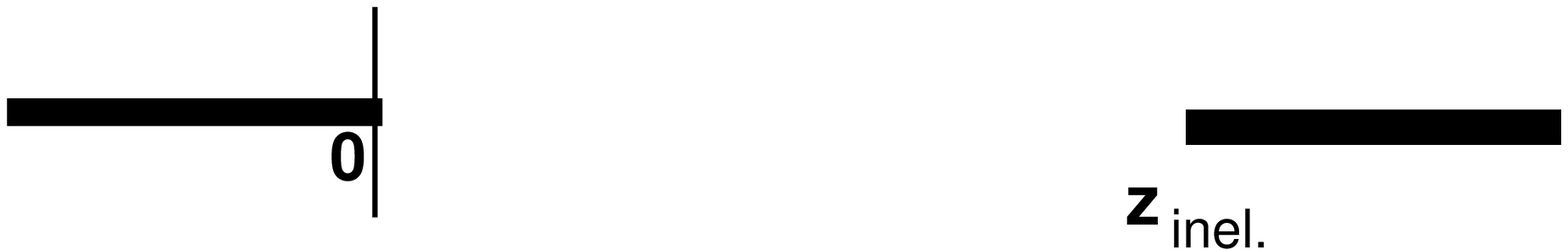}
\vskip.53cm\hspace{8.5cm}\framebox{ cut plane $C_{II}$}

\vskip1cm

\begin{center}
\includegraphics[width=7cm]{CIIIt.eps}\\[3mm]
\framebox{ region $C_{III}$}
\end{center}
\caption{Regions of analyticity in the complex $z$--plane, isospin
  symmetric case. We assume that $t_k$ are holomorphic in the cut plane
  $C_I$. The two-particle irreducible amplitude $t$ is then meromorphic in
  the cut plane $C_{II}$, and the phase--removed form factor $\hat F_c$ is
  holomorphic in the shaded region $C_{III}$.  The symbol $z_{inel}$
  denotes the inelastic threshold $z=16M_{\pi}^2$.}\label{fig:holsymm}
\end{figure}

The partial waves are 
denoted by $t_k$. 
We assume that  these 
i) are holomorphic in the cut plane $C_I$ 
shown in Fig.~\ref{fig:holsymm}, ii) are real on the real axis
$0<s<4M_\pi^2$, and iii) converge
to continuous boundary values $t_k(s)$  as $z\to s$.  The form factors are 
boundary values of functions $F_k(z)$ which are assumed to be
 i) holomorphic
 in the complex $z$-plane,  cut at $z=s$  for $s \geq 4M_{\pi}^2\scs$
ii) real on the real axis for  $s < 4M_{\pi}^2\scs$
iii) nonzero for finite $|z|$.
 Furthermore, we assume that $F_k(z)$ converge to continuous boundary 
values $F_k(s)$  as $z\to s$.

\vskip1cm

We introduce the two--particle irreducible amplitude
\bea\label{eq:tirr}
t(z)=\frac{t_0(z)}{1+i\sigma(z) t_0(z)} \,.
\eea
It is meromorphic  in $C_I$. The poles are due to possible zeros in the 
denominator. In the elastic region, the discontinuity of $t$ vanishes. 
>From the Edge-of-the-Wedge theorem (EWT) it then follows  that $t$ is
meromorphic in the cut plane $C_{II}$ displayed in Fig.~\ref{fig:holsymm}.
Solving  for $t_0$ gives 
\bea
t_0(z)&=&F(z)+i\sigma G(z)\,\,\nnnl
F&=&\frac{t}{1+\sigma^2t^2}\scs G=t F\fs
\eea
The quantities $F,G$ are meromorphic in  $C_{II}$. From this representation, 
it follows that the analytic continuation of $t_0$ to the second Riemann
sheet is
\cite{threshold}
\bea 
t_0^{(2)}=\frac{t_0}{1+2i\sigma t_0}\scs
\eea
which shows  that the zeros of the $S$-matrix on the first Riemann sheet
correspond to poles of the partial waves on the second Riemann sheet, at
the same value of $z$.

Now consider the form factor $F_c$, and define the ratio
\bea\label{eq:R}
R(z)=\frac{F_c(z)}{t_0(z)}\scs
\eea
which is meromorphic in $C_I$. Unitarity and EWT show that the elastic cut
is absent. Therefore, one has the decomposition
\bea
F_c(z)=A(z)+i\sigma B(z)\scs
\eea
with $A,B$ meromorphic in $C_{II}$.  Furthermore, the form factor generates
a pole on the second Riemann sheet as well, at the same place where $t_0$
does.  Finally, the phase defined in Eq.~\eqref{eq:phasessym} coincides
with the isospin zero S--wave $\pi\pi$ phase shift in the elastic interval,
because $R$ is real there. The phase--removed form factor in
Eq.~\eqref{eq:phaseremoved} is holomorphic in the shaded region $C_{III}$
displayed in Fig.~\eqref{fig:holsymm}.

\subsection{Isospin broken case}
\begin{figure}[h]
\hspace{-1mm}\includegraphics[width=6.5cm]{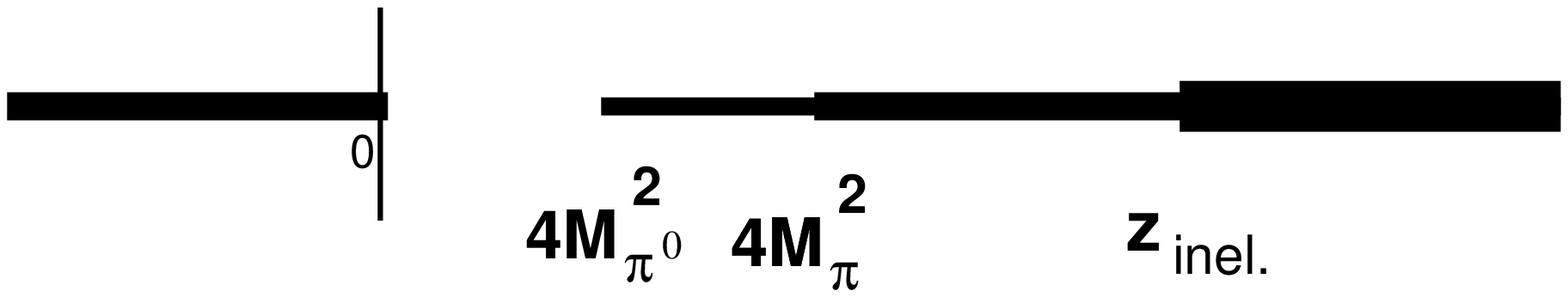} 

\vskip5mm
\hspace{2cm}\framebox{cut plane $D_I$}
\vskip-2.3cm\hspace{7cm}
\includegraphics[width=6.5cm]{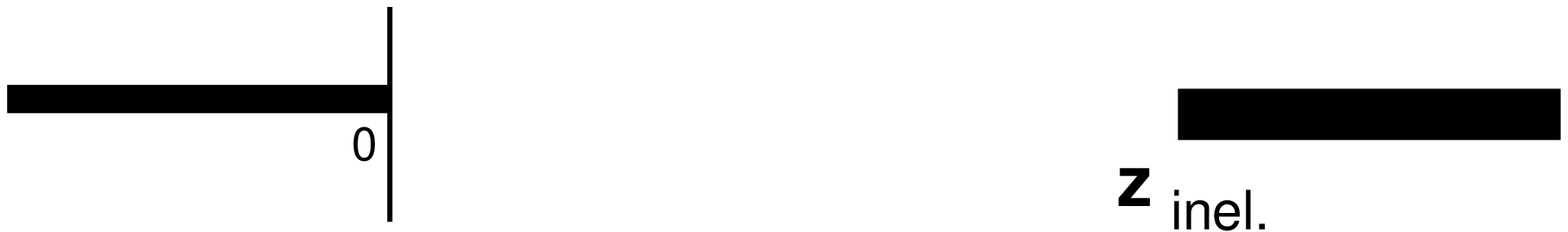}
\vskip.53cm\hspace{8.5cm}\framebox{ cut plane $D_{II}$}

\vskip1cm

\begin{center}
\includegraphics[width=7cm]{DIIIt.eps}\\[3mm]
\framebox{ region $D_{III}$}
\end{center}
\caption{Regions of analyticity in the complex $z$-plane, isopin broken
  case.  We assume that $t_k$ are holomorphic in the cut plane $D_I$. The
  two-particle irreducible amplitude $\hat T$ is meromorphic in the cut
  plane $D_{II}$, and the phase--removed form factor $\hat F_c$ is
  holomorphic in the shaded region $D_{III}$, with the cut indicated.  The
  symbol $z_{inel}$ denotes the inelastic threshold
  $z=16M_{\pi^0}^2$.}\label{fig:holbroken}
\end{figure}

The threshold structure of the form factor displayed in
Eq.~\eqref{eq:threshold_fisobroken} can be obtained in a manner quite
analogously to the previous subsection. One first works out the threshold
behaviour of the scattering amplitudes. We assume that the $h_k$ i) are
holomorphic in the region $D_I$ displayed in Fig.~\ref{fig:holbroken}, ii)
are real on the real axis in the interval $0<s<4M_{\pi^0}^2$, and ii)
converge to continuous boundary values $h_k(s)$ as $z\to s$.  The form
factors $F_k$ are assumed to satisfy the same conditions as above, with
$M_\pi^2\to M_{\pi^0}^2$.

\vskip1cm

We introduce the matrix 
\bea
\bar\rho(z)=\left(\!\!\begin{array}{cc} 2\sigma(z)&0\\0& 
\sigma_0(z)\,\end{array}\!\!\right)\scs
\eea
compare the comments after Eq.~\eqref{eq:sigmas}. Consider the
two--particle irreducible amplitudes
\bea
\hat T=\left(1+iT\bar \rho\right)^{-1}T\scs
\eea
 where the matrix $T$ is defined in Eq.~\eqref{eq:matricesFT}. Unitarity
 and EWT show that the cut $ 4M_{\pi^0}^2<s<16 M_{\pi^0}^2$ is absent for
 $\hat T$, from where we conclude that the matrix $T$ has the threshold
 structure 
\bea
T=A'+i\sigma B'+i\sigma_0 C'+\sigma\sigma_0D'
\eea
with matrices $A',B',C',D'$ whose entries are meromorphic in
$D_{II}$. Next, one forms 
\bea
\left(\!\!\begin{array}{c}
R_c\\ R_0\end{array}\!\!\right)
=
T^{-1}F\scs
\eea
where $F$ is the matrix in Eq.~\eqref{eq:matricesFT}, and uses again
unitarity and EWT to show that $R_{c,0}$ are meromorphic in
$D_{II}$. Solving for $F$ gives
\bea\label{eq:structureFI}
F_k=A_k+i\sigma B_k +i\sigma_0 C_k +\sigma\sigma_0D_k\,;\, k=c,0\scs
\eea
with $A_k,\cdots, D_k$ meromorphic in $D_{II}$. 
Evaluating the phases $\theta_k$ according to Eq.~\eqref{eq:phasesbroken}
shows that these depend on the ratio $R_0/R_c$. On the other hand, this
ratio depends on  $F_c/F_0$, which is unity in the isospin limit, and
different otherwise. In subsection \ref{sec:1loop} we show that, in case of
ChPT, $F_c/F_0$ contains LECs that do not occur in the $\pi\pi$ scattering
amplitude. Hence the phases $\theta_k$ are not determined by the $\pi\pi$
amplitude alone in the case of broken isospin. From
\eqref{eq:phasesbroken}, it is seen that the phase $\theta_c$ does not
vanish at the threshold $s=4M_\pi^2$, because $\sigma_0(4M_\pi^2)\neq 0$. 
Finally, from Eq.~\eqref{eq:structureFI}, one finds for the phase--removed
form factor  the result 
\bea
\hat F_c=(A_c+\sigma\sigma_0 D_c)
\left[1+\left(\frac{\sigma B_c+\sigma_0C_c}{A_c+\sigma\sigma_0 D_c}
\right)^2\right]^\frac{1}{2}\fs
\eea
It is holomorphic in the cut region $D_{III}$ displayed in
Fig. \ref{fig:holbroken}. The cut is due to the factor $\sigma\sigma_0$.

\section{Non-relativistic effective Lagrangian framework}
\label{app:model}

The non-relativistic Lagrangian ${\cal L}={\cal L}_{\pi\pi}
+{\cal L}_{\cal O}$, which is used for calculation of the scalar
form factor, consists of a part describing quartic pion-pion interactions,
and a part corresponding to the interaction of the pion pair with the
external current ${\cal O}(x)$. The framework is described in detail in our
recent works~\cite{cuspwe,cuspwe0,cusprad}  and will not be repeated
here. For example, the Lagrangian ${\cal L}_{\pi\pi}$ is given in Eq. (4)
of Ref.~\cite{cuspwe}. The polynomials $d_i(z)$, introduced in
section~\ref{sec:scalarff}, are given by 
\eq\label{eq:smalld}
32\pi d_{c}&=&C_{+-}+(z-4M_\pi^2)D_{+-}\, ,
\nonumber\\[2mm]
32\pi d_{x}&=&C_{x}+(z-4M_\pi^2)D_{x}\, ,
\nonumber\\[2mm]
32\pi d_{0}&=&C_{00}+(z-4M_{\pi^0}^2)D_{00}\, .
\en
Here we use the same notations for the 4-pion couplings $C,D$ as in 
Refs.~\cite{cuspwe,cuspwe0,cusprad}. The matching of these couplings to the
effective-range expansion parameters in the $\pi\pi$ scattering amplitudes is
described in these references as well.

The part of the Lagrangian which is responsible for the interaction with
the external source takes the form
\eq
{\cal L}_{\cal O}&=&{\cal O}\,\biggl(-f_1\Phi_+^\dagger\Phi_-^\dagger
+f_2(W_\pm\Phi_+^\dagger W_\pm\Phi_-^\dagger
+\nabla\Phi_+^\dagger
\nabla\Phi_-^\dagger-M_\pi^2\Phi_+^\dagger\Phi_-^\dagger) 
\nonumber\\[2mm]
&+&\frac{f_3}{2}\,\Phi_0^\dagger\Phi_0^\dagger
-\frac{f_4}{2}\,(
W_0\Phi_0^\dagger W_0\Phi_0^\dagger
+\nabla\Phi_0^\dagger \nabla\Phi_0^\dagger
-M_{\pi^0}^2\Phi_0^\dagger\Phi_0^\dagger)
\biggr)+\mbox{h.c.}\, .
\en
where $\Phi_\pm,\Phi_0$ denote the non-relativistic charged and neutral 
pion fields, $W_\pm=\sqrt{M_\pi^2-\triangle}$ and 
$W_0=\sqrt{M_{\pi^0}^2-\triangle}$ are the pertinent differential
operators, and $f_1,\cdots f_4$ stand for the various non-relativistic
couplings. The polynomials $f_c,f_0$ which are used in the main text are
defined as 
\eq\label{eq:barf}
f_c=f_1-\frac{1}{2}\,(z-4M_\pi^2)f_2\, ,
\quad\quad f_0=f_3-\frac{1}{2}\,(z-4M_{\pi^0}^2)f_4\, .
\en
In order to calculate the scalar form factors
within the non-relativistic framework, 
one evaluates the transition
amplitude between the { non-relativistic} 
two--pion states and the vacuum in the
presence of the external  source ${\cal O}(x)$. The form factors are
defined by 
\eq\label{eq:defscalar}
\langle \pi^+(p_1)\pi^-(p_2);\mbox{out}|0;\mbox{in}\rangle_{\cal O}^{NR}
&=&i\int d^4x{\rm e}^{i(p_1+p_2)x}{\cal O}(x)\,F_c(s)+O({\cal O}^2)\, ,
\nonumber\\[2mm]
\langle \pi^0(p_1)\pi^0(p_2);\mbox{out}|0;\mbox{in}\rangle_{\cal O}^{NR}
&=&i\int d^4x{\rm e}^{i(p_1+p_2)x}{\cal O}(x)\,F_0(s)+O({\cal O}^2)\, ,\nnnl
\en
where $s=(p_1+p_2)^2$.

The behaviour of both the relativistic and non-relativistic form factors at
small momenta is the same and is displayed in
Eq.~(\ref{eq:threshold_fisobroken}). The polynomials $f_k$ can be
determined by performing a matching of the regular part $A_k$ of the form
factors, evaluated in both theories.


\begin{thebibliography}{99}

\bibitem{weinberg79}
 S.~Weinberg,
  Physica A {\bf 96} (1979) 327.

\bibitem{scattnpb}
 G.~Colangelo, J.~Gasser and H.~Leutwyler,
  Nucl.\ Phys.\  B {\bf 603} (2001) 125
  [arXiv:hep-ph/0103088].

\bibitem{colangelokaon07}
  G.~Colangelo,
  {\em ``Theoretical progress on pi pi scattering lengths and phases,''}
Talk given at Kaon International Conference (KAON'07), Frascati, Italy, 21-25
May 2007,
   PoS {\bf KAON} (2008) 038
  [arXiv:0710.3050 [hep-ph]].



\bibitem{ke4old}
 L.~Rosselet {\it et al.},
  Phys.\ Rev.\  D {\bf 15} (1977) 574;


  S.~Pislak {\it et al.}  [BNL-E865 Collaboration],
  Phys.\ Rev.\ Lett.\  {\bf 87} (2001) 221801
  [arXiv:hep-ex/0106071];

  S.~Pislak {\it et al.}  [BNL-E865 Collaboration],
  Phys.\ Rev.\ D {\bf 67} (2003) 072004
  [arXiv:hep-ex/0301040].


\bibitem{ke4NA48/2}
  J.~R.~Batley {\it et al.}  [NA48/2 Collaboration],
  Eur.\ Phys.\ J.\  C {\bf 54} (2008) 411.

\bibitem{blochkaon07}
B.~Bloch-Devaux,
 {\it Recent results from NA48/2 on K(e4) decays and interpretation in term of $\pi
 \pi$ scattering lengths},
  PoS {\bf KAON} (2008) 035.

\bibitem{blochanacapri}
B.~Bloch-Devaux, {\it Results from NA48/2 $K_{e4}$ decays: Form Factors and $\pi\pi$ scattering lengths},
talk given at: FlaviaNet Kaon Workshop, Anacapri, Italy, June 12-14, 2008.

\bibitem{talksna48}
For talks provided by members of the NA48/2 collaboration, see\newline
http://na48.web.cern.ch/NA48/Welcome/images/talks.html



\bibitem{DIRAC}
B.~Adeva {\it et al.}  [DIRAC Collaboration],
  J.\ Phys.\ G {\bf 30} (2004) 1929
  [arXiv:hep-ex/0409053];

 B.~Adeva {\it et al.}  [DIRAC Collaboration],
  Phys.\ Lett.\  B {\bf 619} (2005) 50
  [arXiv:hep-ex/0504044].

\bibitem{k3pNA48kTeV}
  J.~R.~Batley {\it et al.}  [NA48/2 Collaboration],
  Phys.\ Lett.\  B {\bf 633} (2006) 173
  [arXiv:hep-ex/0511056];


E. Abouzaid et al. [The KTeV Collaboration],
  arXiv:0806.3535 [hep-ex].





\bibitem{photos}
  E.~Barberio, B.~van Eijk and Z.~Was,
  Comput.\ Phys.\ Commun.\  {\bf 66} (1991) 115;

 E.~Barberio and Z.~Was,
  Comput.\ Phys.\ Commun.\  {\bf 79} (1994) 291;

 
G.~Nanava and Z.~Was,
  Eur.\ Phys.\ J.\  C {\bf 51} (2007) 569
  [arXiv:hep-ph/0607019].


\bibitem{internalnote}
J.~Gasser and A.~Rusetsky, {\it Isospin violations in $K_{e4}$ decays}, Internal 
note to the NA48/2 collaboration, March 2007;

  J.~Gasser,
  PoS {\bf KAON} (2008) 033
  [arXiv:0710.3048 [hep-ph]];

G.~Colangelo, Ref.~\cite{colangelokaon07}.

\bibitem{tarasovke4}
 S.~R.~Gevorkyan, A.~N.~Sissakian, A.~V.~Tarasov,
 H.~T.~Torosyan and O.~O.~Voskresenskaya,
  arXiv:0704.2675 [hep-ph].

\bibitem{gevorkyanke4}
  S.~R.~Gevorkyan, A.~N.~Sissakian, A.~V.~Tarasov, H.~T.~Torosyan and O.~O.~Voskresenskaya,
  arXiv:0711.4618 [hep-ph].

\bibitem{descotesanacapri}
 S.~Descotes-Genon, 
{\it $\pi\pi$ scattering: isospin breaking corrections},
talk given at: FlaviaNet Kaon Workshop, Anacapri, Italy, June 12-14, 2008.


\bibitem{cuplov}
 V.~Cuplov and A.~Nehme,
  arXiv:hep-ph/0311274;

 A.~Nehme,
  Phys.\ Rev.\  D {\bf 69} (2004) 094012
  [arXiv:hep-ph/0402007];
 
 A.~Nehme,
  Eur.\ Phys.\ J.\  C {\bf 40} (2005) 367
  [arXiv:hep-ph/0408104].


\bibitem{cabibbomaksymovicz}
N.~Cabibbo and A.~Maksymovicz, Phys. Rev. 137 (1965) B438; erratum Phys. Rev. 168 (1968) 1926.

\bibitem{partialwaveexpansion}
F.~A.~Berends, A.~Donnachie and G.~C.~Oades, Phys. Lett. 26B (1967) 109; Phys. Rev. 171 (1968) 1457.

\bibitem{kl4oneloop}
 J.~Bijnens,
  Nucl.\ Phys.\  B {\bf 337} (1990) 635 ;

  C.~Riggenbach, J.~Gasser, J.~F.~Donoghue and B.~R.~Holstein,
  Phys.\ Rev.\  D {\bf 43} (1991) 127.

\bibitem{chiralp4}
   R.~Urech,
  Nucl.\ Phys.\  B {\bf 433} (1995) 234
  [arXiv:hep-ph/9405341];
 
 H.~Neufeld and H.~Rupertsberger,
  Z.\ Phys.\  C {\bf 71} (1996) 131
  [arXiv:hep-ph/9506448];

 U.-G.~Mei{\ss}ner, G.~Muller and S.~Steininger,
  Phys.\ Lett.\  B {\bf 406} (1997) 154
  [Erratum-ibid.\  B {\bf 407} (1997) 454]
  [arXiv:hep-ph/9704377];

  M.~Knecht and R.~Urech,
  Nucl.\ Phys.\  B {\bf 519} (1998) 329
  [arXiv:hep-ph/9709348].

\bibitem{AnnalsNPB}
  J.~Gasser and H.~Leutwyler,
  Annals Phys.\  {\bf 158} (1984) 142;

 J.~Gasser and H.~Leutwyler,
  Nucl.\ Phys.\  B {\bf 250}, 465 (1985).


\bibitem{internalknecht}
M.~Knecht, {\it Isospin breaking in the phases of two--pion states}, Internal note to the NA48/2 collaboration, 
June 2007.



\bibitem{chptlept}
  M.~Knecht, H.~Neufeld, H.~Rupertsberger and P.~Talavera,
  Eur.\ Phys.\ J.\ C {\bf 12} (2000) 469 
  [arXiv:hep-ph/9909284].

\bibitem{threshold}
W.~Zimmermann, Nuov. \ Cim. \  {\bf 21} (1961) 249, and references cited therein.


\bibitem{cuspwe}
  G.~Colangelo, J.~Gasser, B.~Kubis and A.~Rusetsky,
  Phys.\ Lett.\  B {\bf 638} (2006) 187
  [arXiv:hep-ph/0604084].

\bibitem{cuspwe0}
  M.~Bissegger, A.~Fuhrer, J.~Gasser, B.~Kubis and A.~Rusetsky,
  Phys.\ Lett.\  B {\bf 659} (2008) 576
  [arXiv:0710.4456 [hep-ph]].



\bibitem{cusprad}
  M.~Bissegger, A.~Fuhrer, J.~Gasser, B.~Kubis and A.~Rusetsky,\\{}
  arXiv:0807.0515 [hep-ph].


\bibitem{physrep}
  J.~Gasser, V.~E.~Lyubovitskij and A.~Rusetsky,
  Phys.\ Rept.\  {\bf 456} (2008) 167
  [arXiv:0711.3522 [hep-ph]].

\bibitem{Kubis:1999db}
  B.~Kubis and U.-G.~Mei{\ss}ner,
  Nucl.\ Phys.\  A {\bf 671} (2000) 332
  [Erratum-ibid.\  A {\bf 692} (2001) 647]
  [arXiv:hep-ph/9908261].


\bibitem{Colangelo:2001sp}
  G.~Colangelo, J.~Gasser and H.~Leutwyler,
  Phys.\ Rev.\ Lett.\  {\bf 86} (2001) 5008
  [arXiv:hep-ph/0103063].



\bibitem{Ananthanarayan:2000ht}
  B.~Ananthanarayan, G.~Colangelo, J.~Gasser and H.~Leutwyler,
  Phys.\ Rept.\  {\bf 353} (2001) 207
  [arXiv:hep-ph/0005297].


\bibitem{Colangelo:2003hf}
  G.~Colangelo and S.~Durr,
  Eur.\ Phys.\ J.\  C {\bf 33} (2004) 543
  [arXiv:hep-lat/0311023].

\bibitem{blochprivate}
B.~Bloch-Devaux and P.~Tru\"ol, private communication.



\end{thebibliography}
\end{document}